\documentclass[superscriptaddress,aps,prd,showpacs,preprint]{revtex4-2}

\usepackage{eurosym}
\usepackage{graphicx}
\usepackage{amsmath}
\usepackage{mathrsfs}
\usepackage{bm}
\usepackage{color}
\usepackage{amsmath}
\usepackage{amssymb}
\usepackage{multirow}
\usepackage{array}
\usepackage[caption=false]{subfig}

\def\be{\begin{equation}}
\def\ee{\end{equation}}
\def\bea{\begin{eqnarray}}
\def\eea{\end{eqnarray}}

\allowdisplaybreaks

\begin{document}

\title{Radiation Spectra and Effective Temperatures\\in Higher-Dimensional Charged-de Sitter Black Holes}

\author{Supakchai Ponglertsakul}
\email{supakchai.p@gmail.com}
\affiliation{Division of Physics and Semiconductor Science, Dongguk University, Seoul 04620,
Republic of Korea}
\author{Bogeun Gwak}
\email{rasenis@dgu.ac.kr}
\affiliation{Division of Physics and Semiconductor Science, Dongguk University, Seoul 04620,
Republic of Korea}

\begin{abstract}
In this study, a scalar field propagating in a higher-dimensional Reissner-Nordstr\"om-de Sitter black hole is investigated. Scalar fields are assumed to have non-minimal coupling to the brane or bulk scalar curvature. Five different {definitions} of black hole temperatures are discussed: temperature based on surface gravity, Bousso-Hawking temperature, and three effective temperatures. The greybody factors of minimally and non-minimally coupled scalar fields on the brane and in the bulk are examined under the effect of particle and spacetime properties. The energy emission spectra of black holes are determined at various temperatures for both the brane and bulk channels. The energy emission rates at the Bousso-Hawking temperature are found to be dominant over those at other temperatures. The energy emission curves are suppressed by the presence of coupling parameters. Finally, the bulk-over-brane emission ratios are calculated. Notably, bulk dominance becomes possible for a certain definition of the temperatures and regime of the cosmological constant.
\end{abstract}

\maketitle

\newpage

\section{Introduction}

The existence of black holes is predicted using Einstein's general theory of relativity. Black holes are described as solutions to Einstein's field equation. The first exact black hole solution discovered by Karl Schwarzschild in 1916 serves as an analytic tool for probing the properties of gravity under extreme conditions. Four-dimensional black holes have been studied both intensively and extensively. The existence and physical properties of black holes have been thoroughly explored. 

In contrast, higher-dimensional theories of gravity have received significant attention as a possible framework for the unification of four fundamental interactions. These extra-dimensional theories suggest the possibility of the existence of higher-dimensional black objects \cite{Emparan:2008eg}. The first higher-dimensional spherically symmetric black hole is known as the Tangherlini solution \cite{Tanghherlini:1963}. It generalizes the Schwarzschild solution of general relativity to higher dimensions in the presence of the cosmological constant. In addition to black holes and gravity, cosmology and particle physics have also been explored in the light of higher-dimensional theories \cite{Randall:1999ee,RUBAKOV1983136,Aghababaie:2003wz,Raychaudhuri:2016kth}. 

One aspect of black hole physics that has received significant attention from physicists is Hawking radiation which is the emission of particles from a black hole owing to the quantum mechanical effect near an event horizon \cite{Hawking:1974sw}. In the spirit of higher-dimensional theories, the emission of Hawking radiation spectra from higher-dimensional black holes has been explored by several authors, in spherically symmetric or axially symmetric setups \cite{Casanova:2005id,Kanti:2008eq,Kanti:2012jh,Kanti:2014vsa,Winstanley:2007hj,Harris:2003eg,Cornell:2005ux,Chen:2008ra,Jung:2005pk,Duffy:2005ns,Casals:2005sa,Casals:2006xp,Chen:2007pu,Liu:2010zzo}. While several studies have investigated the Hawking radiation of a higher-dimensional Schwarzschild black hole, only a few studies have focused on their de Sitter (dS) counterparts. For a higher-dimensional Schwarzschild-dS black hole, minimally coupled scalar radiations on the brane and in the bulk were studied in \cite{Kanti:2005ja}. In \cite{Harmark:2007jy}, an analytic study of the transmission amplitude or greybody factor under the same conditions was performed. In addition to the scalar field, the Hawking emission of fields with arbitrary spins \cite{Wu:2008rb} and greybody factors of the fermionic field \cite{Sporea:2015hla} on the Scwharzschild-dS black hole have been investigated. Moreover, in the non-minimally coupled scalar field sector, the greybody factor was studied in \cite{Crispino:2013pya} {for} the four-dimensional Schwarzschild-dS {spacetime} and in \cite{Zhang:2017yfu,Zhang:2020qam,Li:2019bwg} {for} Gauss-Bonnet black holes. This was later extended to a higher dimension by Kanti et al. \cite{Kanti:2014dxa}, where the scalar field existed either on the brane or in the bulk. Later, the Hawking radiation spectra and greybody factor for non-minimally coupled scalar fields in the $D$-dimensional Schwarzschild-dS black hole were computed in \cite{Pappas:2016ovo}.

Furthermore, many studies have focused on fields and particles propagating from charged black holes. For a spherically symmetric black hole, Hawking radiation can be understood based on the tunneling phenomenon \cite{Sarkar:2007sx}. The emission of Hawking radiation for the Reissner-Nordstr\"om (RN) black hole was determined using the tunneling method \cite{Jiang:2012dg}. Hawking radiation and fermion tunneling have been studied in and beyond the semiclassical limits for higher-dimensional RN black holes \cite{Yang:2016bqm}. In addition, the absorption cross section of a massive scalar field propagating from a charged black hole has been numerically computed at intermediate frequencies \cite{Benone:2014qaa}. A series of studies has been undertaken to examine the absorption and emission spectra of a higher-dimensional RN black hole for brane and bulk scalars \cite{Jung:2005mr,Konoplya:2008au,Konoplya:2014lha}, Dirac fermions \cite{Jung:2005sw}, and electromagnetic waves \cite{Crispino:2010fd}. The bounds of the greybody factor for the RN black hole have been determined using transfer matrices \cite{Ngampitipan:2013sf}. Few researchers have shown interest in the emission of Hawking radiation spectra for charged black holes in non-asymptotically flat spacetime. In \cite{Lai:2008zzb}, fermion tunneling from the Reissner-Nordstr\"om-anti-de Sitter (RN-AdS) black hole was considered. In addition, the greybody factor of non-minimally coupled scalar fields in RN-dS was discussed in \cite{Ahmed:2016lou}. Using the tunneling method, Wu and Jian calculated Hawking radiation-charged particles from higher-dimensional RN-dS black holes \cite{Wu:2006nj}.

In the standard black hole thermodynamics, the temperature of the black hole is based on its surface gravity associated with a horizon. A problem arises when the black hole possesses a positive cosmological constant. Such a spacetime has an upper boundary, referred to as the cosmological horizon. Therefore, an observer living in the region bounded by an event horizon and a cosmological horizon cannot be in thermodynamic equilibrium. Heat always flows from the hotter (event) horizon to the colder (cosmological) horizon. Moreover, defining black hole parameters in the dS spacetime is subtle, as the notion of these parameters is securely defined {only on} an asymptotically flat spacetime. In \cite{Bousso:1996au}, Bousso and Hawking proposed a normalized black hole temperature at which the value of the cosmological constant is assumed to be small; thus, the two horizons are located far apart. Therefore, the two horizons can be treated as two independent thermodynamic systems. The effective temperature of the dS black hole was proposed \cite{Shankaranarayanan:2003ya,Li:2016zca,Kubiznak:2016qmn} to consider a large cosmological constant scenario. In \cite{Kanti:2017ubd}, the Hawking emission spectra of minimally and non-minimally coupled scalar fields on the brane and in the bulk for a higher-dimensional Schwarzschild-dS black hole were examined under five different {definitions} of black hole temperatures, namely, the temperature based on surface gravity, Bousso-Hawking temperature, and three effective temperatures. The results indicate that different temperatures can lead to different outcomes in terms of the domination of the brane or bulk emission channel.

Herein, we extend the study of Kanti and Pappas \cite{Kanti:2017ubd} by considering the effect of the charge of the black hole on energy emission spectra and comparing the power spectra at five different temperatures. The remainder of this paper is organized as follows. In section~\ref{sect:spacetime}, a higher-dimensional RN-dS black hole is discussed. A selective black hole phase space is also explored. In Section~\ref{sect:BHtemp}, we define five different black hole temperatures. In section~\ref{sect:greybody}, we present the equation of motion for a non-minimally coupled scalar field. The greybody factor is also derived and numerically computed for brane and bulk scalar fields. In section~\ref{sect:EER}, we calculate and compare the energy emission rates (EERs) obtained under the effect of the five definition temperatures. The bulk-over-brane total energy emission ratio is compared in Section~\ref{sect:Bulkoverbrane}. We present our conclusions in Section~\ref{sect:conclude}.

\section{Spacetime background}\label{sect:spacetime}

The action describing the higher-dimensional Einstein-Maxwell theory with a cosmological constant $\Lambda$ is defined as 
\begin{align}
S &= \int d^{n+4}\sqrt{-g}\left[\frac{\mathcal{R}}{2} - \Lambda - \frac{1}{4} F_{\mu\nu}F^{\mu\nu}\right], 
\end{align}
where $\mathcal{R}$ and $F_{\mu\nu}$ are the Ricci scalar and Maxwell tensor, respectively. When we vary this action with respect to the metric tensor, we obtain the Einstein field equation
\begin{align}
R_{\mu\nu} - \frac{1}{2}g_{\mu\nu}\mathcal{R} + \Lambda g_{\mu\nu} &= F_{\mu\rho}F_{\nu}^{\rho} - \frac{1}{4}F_{\sigma\gamma}F^{\sigma\gamma}. 
\end{align}
Because this theory admits a static spherically symmetric background, its line element is expressed as
\begin{align}
ds^2 = -fdt^2 + f^{-1}dr^2 + r^2d\Omega^2_{n+2}, \label{bulkmetric}
\end{align}
where the metric of the $(n+2)$ sphere is 
\begin{align}
d\Omega^2_{n+2} &= d\theta_1^2 + \sum_{i=2}^{n+2}\left[\prod_{j=1}^{i-1}\left(\sin^2\theta_{j}\right)\right]d\theta_i^2.
\end{align}
The metric function is explicitly defined as
\begin{align}
f &= 1 - \frac{2M}{r^{n+1}} + \frac{Q^2}{r^{2(n+1)}} - r^2\bar{\Lambda}. \label{metricfunction}
\end{align}
The mass and charge of the black hole are denoted as $M$ and $Q$, respectively. The cosmological constant is $\bar{\Lambda}\equiv \frac{2\Lambda}{(n+2)(n+3)}$. The real positive roots determine the locations of the horizons of the black hole. A charged dS black hole typically has three horizons, namely, the Cauchy $(r_0)$, event $(r_h)$, and cosmological horizons $(r_c)$, where $r_0<r_h<r_c$.

The mass of the black hole, $M$, can be related to other background parameters $Q$ and $\Lambda$. Considering $f(r_h)=0$, we get
\begin{align}
M &= \frac{1}{2r_h^{n+1}}\left(Q^2 + r_h^{2n+2}\left(1-r_h^2\bar{\Lambda}\right)\right). \label{Msub}
\end{align}
To ensure the presence of the event horizon, we require $f(r_h)=0$ and $f'(r_h)\geq 0$. With fixed $r_h=1$ and $Q,\Lambda\geq0$, we obtain the following condition
\begin{align}
\bar{\Lambda} &\leq \frac{(n+1)(1-Q^2)}{n+3}. 
\end{align}
For this study, the event horizon of the black hole was fixed at $1$.
\begin{figure}[h]
\includegraphics[width=0.5\textwidth]{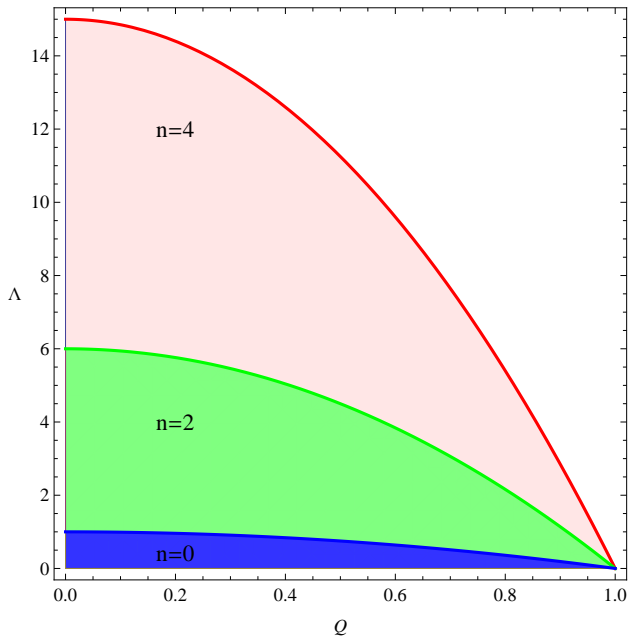}
\caption{Parameter space of higher-dimensional RN-dS black hole with $r_h=1$. The shaded regions denote the area where the black hole has three horizons.} 
\label{phasespace}
\end{figure}
Therefore, the charges on the black holes vary within the range of $0 \leq Q < 1$. Fig.~\ref{phasespace} illustrates the parameter space of  higher-dimensional RN-dS black holes for various spacetime dimensions $n$. Black holes with three horizons, namely, $r_0<r_h<r_c$, can be identified in the colored area in the plot. In addition, the Nairai limit $(r_h\to r_c)$ \cite{Nariai} is described by the {\it{extremal curve}} located at the boundaries of these plots. It is evident that when the charge of the black hole increases, black holes with three horizons exist, with a small value of $\Lambda$. In addition, the allowed value of $\Lambda$ (i.e., the colored area) increases when the number of extra spacetime dimensions increases.

In this study, we consider the EER of scalar propagation in the brane and bulk spacetimes. Any nonstandard model of particles may travel in bulk, as described by the background metric (\ref{metricfunction}). In contrast, ordinary particles are constrained to propagate only on a four-dimensional brane, where an observer exists. This four-dimensional brane is described by the following gravitational background metric \cite{Kanti:2017ubd}.
\begin{align}
ds^2 = -fdt^2 + f^{-1}dr^2 + r^2d\Omega^2. \label{branemetric}
\end{align}
On the $4D$ brane, the extra-hyper-angular $d\theta_3=...=d\theta_{n+2}$ is fixed to zero such that an ordinary particle moves only in four dimensions. {However, it must be emphasized that the free particle living on the $4D$ brane will move differently from those living in the usual four-dimensional spacetime. This is because metric function $f$ on the brane decays faster than the standard $f$ function in four dimensions; therefore, their gravitational potentials are different.} Note that the metric function preserves the form expressed in (\ref{metricfunction}). Therefore, the horizon structure and parameter space analysis discussed earlier can also be applied to the brane scenario.

\section{Black hole temperature}\label{sect:BHtemp}

In this study, we investigated the EER of a (non-)minimally coupled scalar field on bulk and brane spacetimes. Note that the EER formula depends on the temperature of the black hole. While this temperature has been widely studied by many researchers, most studies have been performed on asymptotically flat black holes. For the black holes in the dS spacetime, their temperature requires careful analysis. Therefore, in this section, we explore the various definitions of temperatures of non-asymptotically flat black holes. 

The spacetime metrics (\ref{bulkmetric}) and (\ref{branemetric}) have time-translational symmetry. They admit a time-like killing vector $\xi = \frac{\partial}{\partial t} $. {For} the spherically symmetric {background}, the surface gravity and temperature of the black hole are expressed as
\begin{align}
\kappa_i &= \frac{f'}{2}\bigg\rvert_{r=r_i}, \\
T_i &= \frac{\kappa_i}{2\pi},
\end{align} 
where surface gravity can be evaluated at the Cauchy, event, and cosmological horizons, that is, $i=\{0,h,c\}$. Therefore, the temperatures of the higher-dimensional RN-dS black holes at the event and cosmological horizons are conventionally expressed as
\begin{align}
T_h &= \frac{1}{4\pi r_h^{2n+3}}\left(r_h^{2n+2}\left((n+1)-(n+3)r_h^2\bar{\Lambda}\right)-(n+1)Q^2\right), \label{TH} \\
T_c &= -\frac{1}{4\pi r_c^{2n+3}}\left(r_c^{2n+2}\left((n+1)-(n+3)r_c^2\bar{\Lambda}\right)-(n+1)Q^2\right). \label{TC}
\end{align}
When $\kappa_c <0$, an additional minus sign is included in $T_c$. In the presence of the cosmological horizon, the thermodynamics of black holes in the dS spacetime becomes more complicated than that in the asymptotically flat spacetime, because each horizon has its own temperature. When $T_h>T_c$, there is a continuous flow of thermal energy from the event horizon to the cosmological horizon, and the observers in this region are not in thermodynamic equilibrium. The temperatures (\ref{TH}) and (\ref{TC}) of the black hole are determined under the assumption that the cosmological horizon is located relatively far away from the event horizons. Therefore, the temperature at each horizon can be treated as its own independent thermodynamic state. This assumption is valid only for a small value of $\bar{\Lambda}$. 

To improve the notion of black hole temperature, Bousso and Hawking proposed a normalized temperature of the black hole \cite{Bousso:1996au} with the following formula
\begin{align}
T_{BH} &= \frac{T_h}{\sqrt{f(r_m)}}, \label{TBH}
\end{align}
where $\frac{1}{\sqrt{f(r_m)}}$ is the normalization constant of the killing vector, and $r_m$ is the location of the global maximum of $f$. In the absence of charge $Q$, $r_m$ can be easily determined \cite{Kanti:2017ubd} using $f'(r_m)=0$. In general, $r_m$ must be chosen such that it is located inside the causally connected region $r_h<r_m<r_c$; otherwise, the temperature (\ref{TBH}) becomes a complex number.  

Recently, another definition of the temperature of black holes has attracted significant attention. The \textit{effective temperature} of black holes attempts to unify $T_h$ and $T_c$ into one formula. In the conventional analysis of black hole thermodynamics, the mass and cosmological constant of the black hole are often treated as enthalpy and pressure, respectively, when formulating the first law of black holes. In addition, the total entropy is assumed to be the sum of both horizons, $S=S_h+S_c$. In this framework, the effective temperature has the form \cite{Kanti:2017ubd,Shankaranarayanan:2003ya,Kubiznak:2016qmn}
\begin{align}
T_{eff-} &= \frac{T_h T_c}{T_h-T_c}.
\end{align}
This can be expressed explicitly as 
\begin{align}
T_{eff-} &= \frac{\left((n+1)Q^2 + r_h^{2n+2}\left((n+3)r_h^2\bar{\Lambda}-(n+1)\right)\right) \left((n+1)Q^2 + r_c^{2n+2}\left((n+3)r_c^2\bar{\Lambda}-(n+1)\right)\right)}{4\pi\left((n+1)(r_h^{2n+3}+r_c^{2n+3})Q^2 + (r_hr_c)^{2n+2}(r_h+r_c)((n+3)r_hr_c\bar{\Lambda}-(n+1))\right)}. \label{Teffm}
\end{align}
Parameter $T_{eff-}$ reduces to the temperature of the cosmological horizon $T_c$ in the limit $r_h\to 0$. In contrast, when the system becomes pressure-less ($\bar{\Lambda}=0$ or equivalently $r_c\to\infty$), the effective temperature vanishes. This indicates that $T_{eff-}$ is not valid in the absence of a cosmological constant \cite{Kanti:2017ubd}. Furthermore, this effective temperature sometimes yields negative results and is ill-defined at a critical point where the temperature becomes infinitely high. This aspect will be revisited later, when the various definitions of temperature are compared.

To resolve the issue of the unphysical result of $T_{eff-}$, a new effective temperature was proposed by \cite{Shankaranarayanan:2003ya,Kubiznak:2016qmn}. In contrast with $T_{eff-}$, the total entropy was expressed as the difference between the entropies of the two horizons, $S=S_c-S_h$. An {\it{ad hoc}} formula for the temperature is 
\begin{align}
T_{eff+} &= \frac{T_h T_c}{T_h+T_c}, \nonumber \\
&= \frac{\left((n+1)Q^2 + r_h^{2n+2}\left((n+3)r_h^2\bar{\Lambda}-(n+1)\right)\right) \left((n+1)Q^2 + r_c^{2n+2}\left((n+3)r_c^2\bar{\Lambda}-(n+1)\right)\right)}{4\pi\left((n+1)(r_c^{2n+3}-r_h^{2n+3})Q^2 - (r_hr_c)^{2n+2}(r_c-r_h)((n+3)r_hr_c\bar{\Lambda}+(n+1))\right)}. \label{Teffp}
\end{align}
$T_{eff+}$ is similar to $T_{eff-}$, that is, $T_{eff+}\to T_c$ as $r_h\to0$ and $T_{eff+}\to 0$ as $r_c\to\infty$. However, at the critical point, $T_{eff+}$ vanishes instead of exhibiting an infinite jump, as in the case of $T_{eff-}$. This is because the numerator in (\ref{Teffp}) becomes zero faster than its denominator.

Finally, a new form of effective temperature was proposed by Kanti and Pappas \cite{Kanti:2017ubd}. Motivated by the effective temperature discussed above, the new effective temperature is given by
\begin{align}
T_{effBH} &= \frac{T_{BH} T_c}{T_{BH}-T_c}. \label{TeffBH}
\end{align}
Because its explicit form is rather lengthy and complex, we decided not to explicitly display it here. This definition of the temperature of the black hole inherits several features from the aforementioned formulae. First, in the limit $T_h\to0$, $T_{effBH}$ reduces to $T_c$ and becomes zero when $\bar{\Lambda}\to 0$. Second, the conventional temperature $T_h$ is replaced by the normalized temperature $T_{BH}$. Third, in the absence of the charge of the black hole, $T_{effBH}$ is found to be zero at the critical point \cite{Kanti:2017ubd}. As we approach the critical point, the numerator in $T_{effBH}$ falls to zero faster than its denominator \cite{Kanti:2017ubd}. However, as will be seen below, this is not always the case when $Q\neq 0$.

\begin{figure}[h]
\subfloat[]{\includegraphics[width=3in]{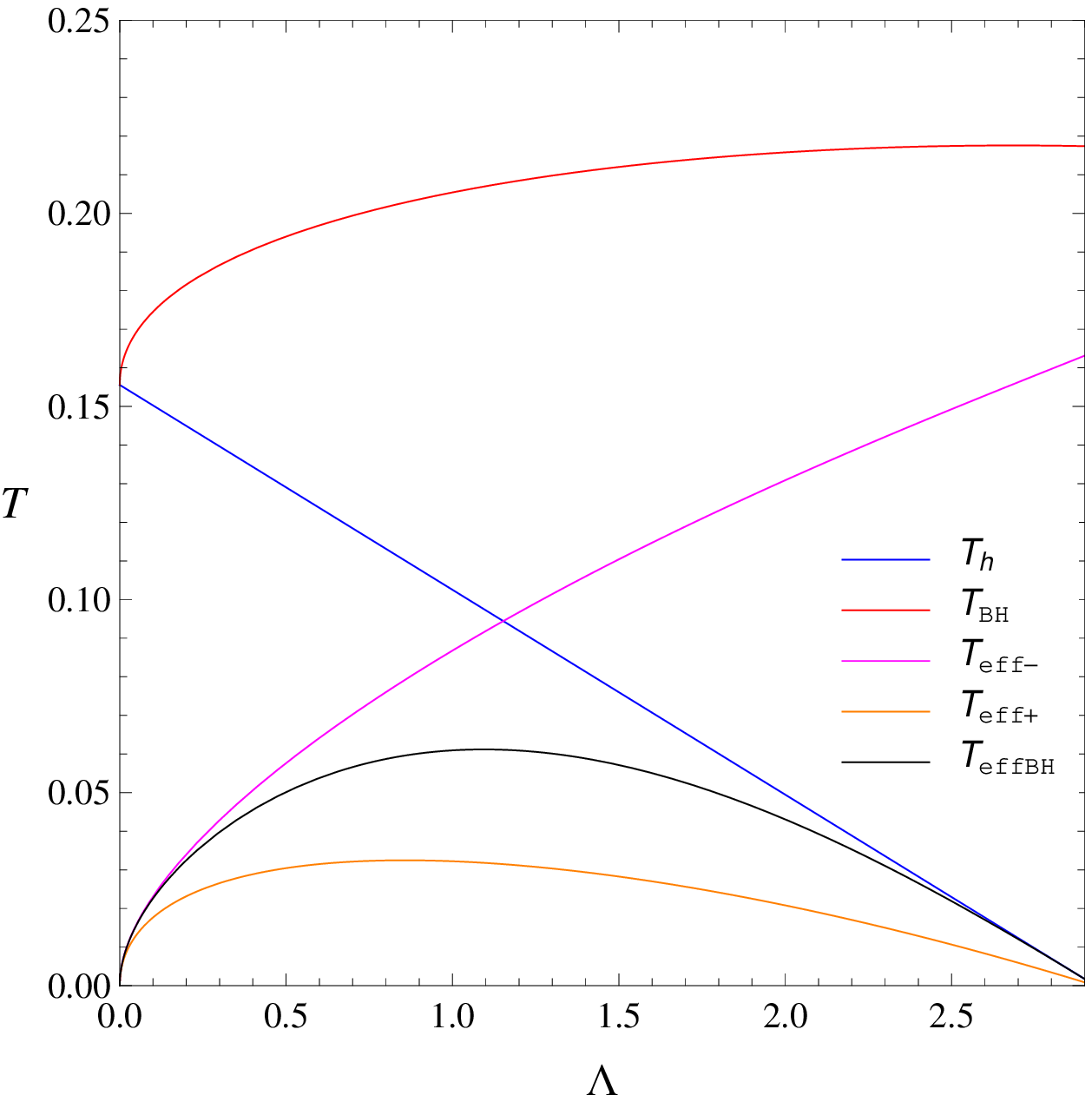} }
\subfloat[]{\includegraphics[width=3in]{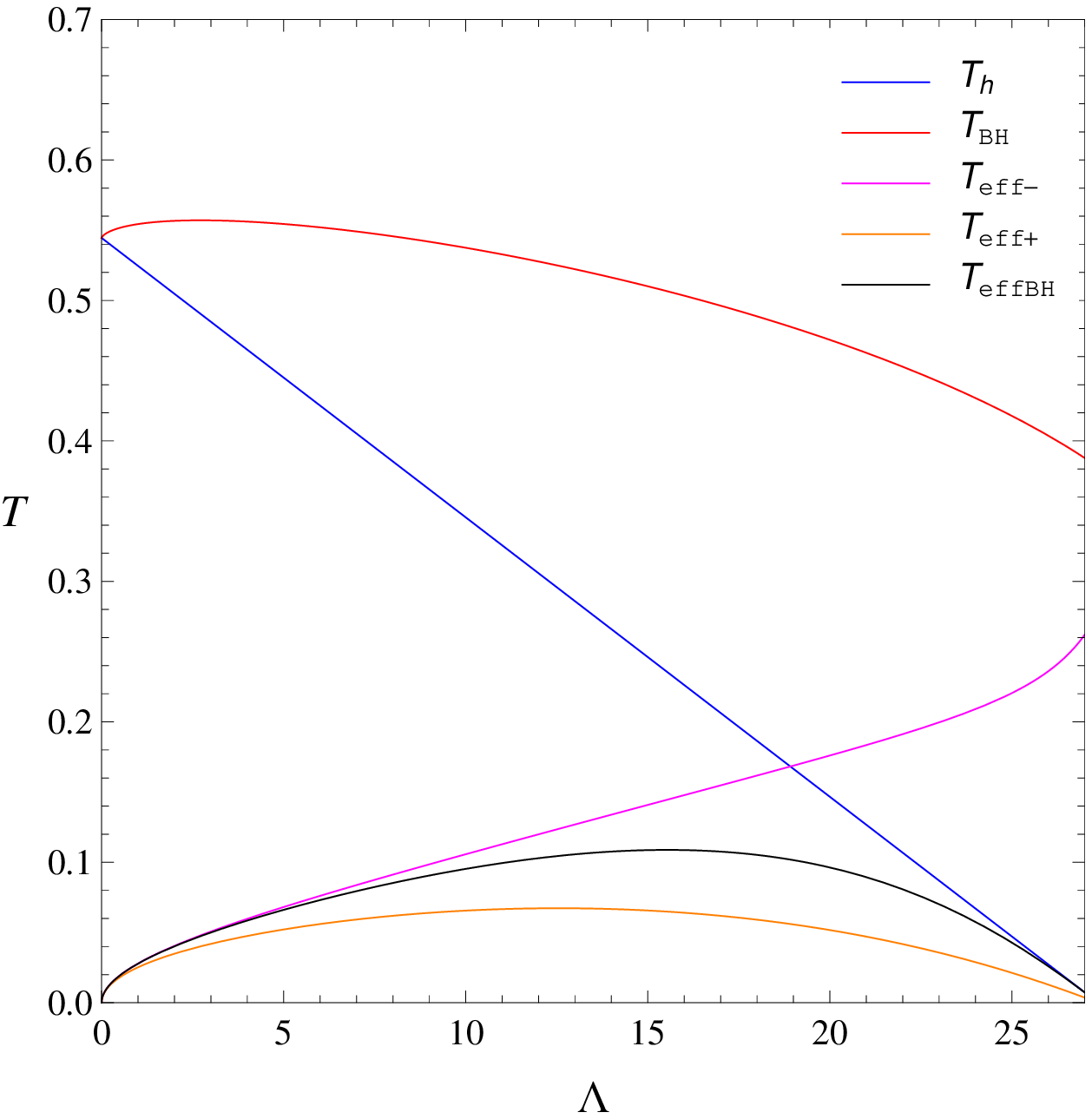}} 
\caption{Comparisons of temperatures of higher-dimensional RN-dS black hole with $Q=0.15$, (a) $n=1$, (b) $n=6$.} 
\label{FIGtemp1}
\end{figure}

\begin{figure}[h]

\subfloat[]{\includegraphics[width=0.45\textwidth]{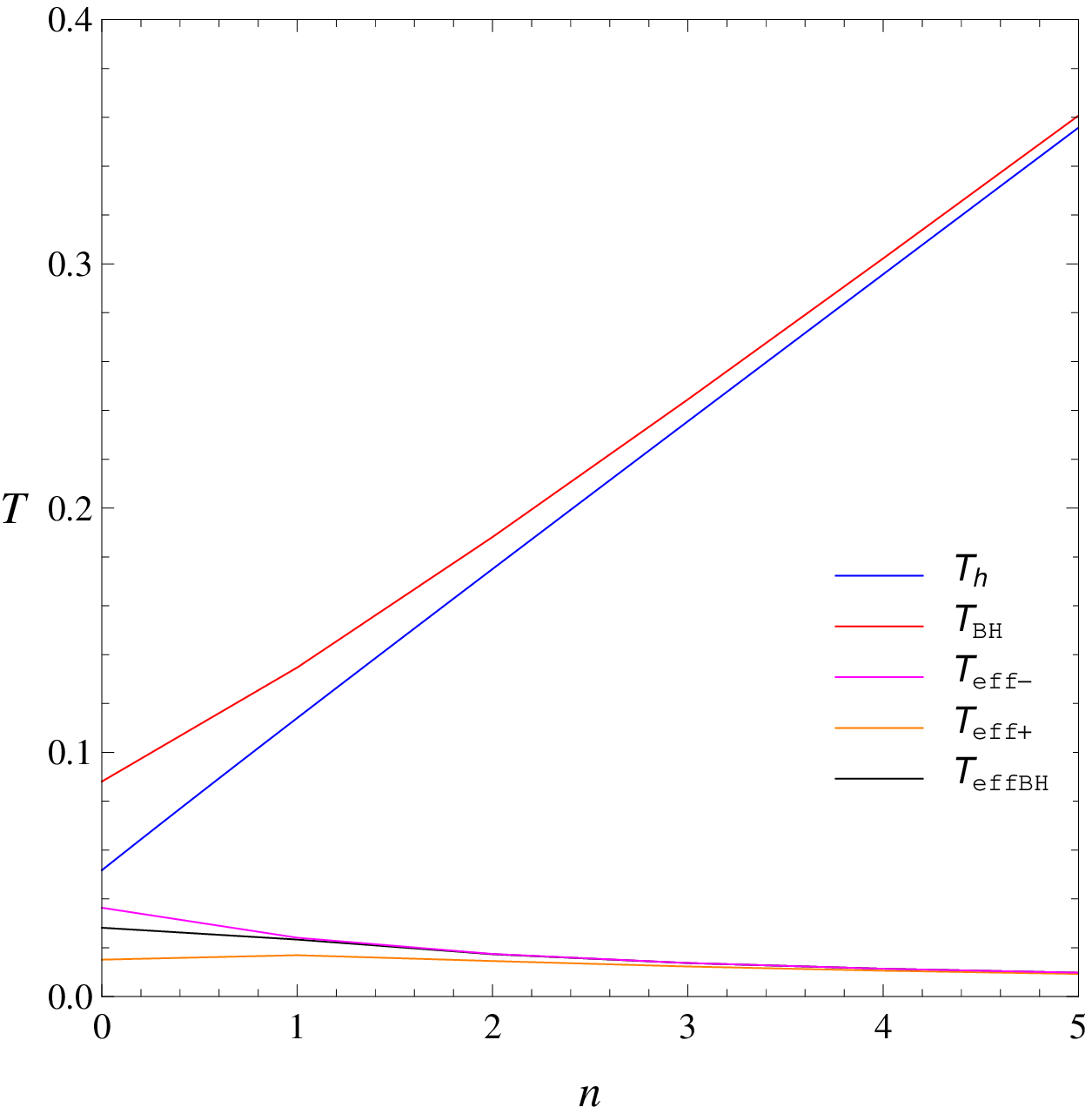}}
\subfloat[]{\includegraphics[width=0.45\textwidth]{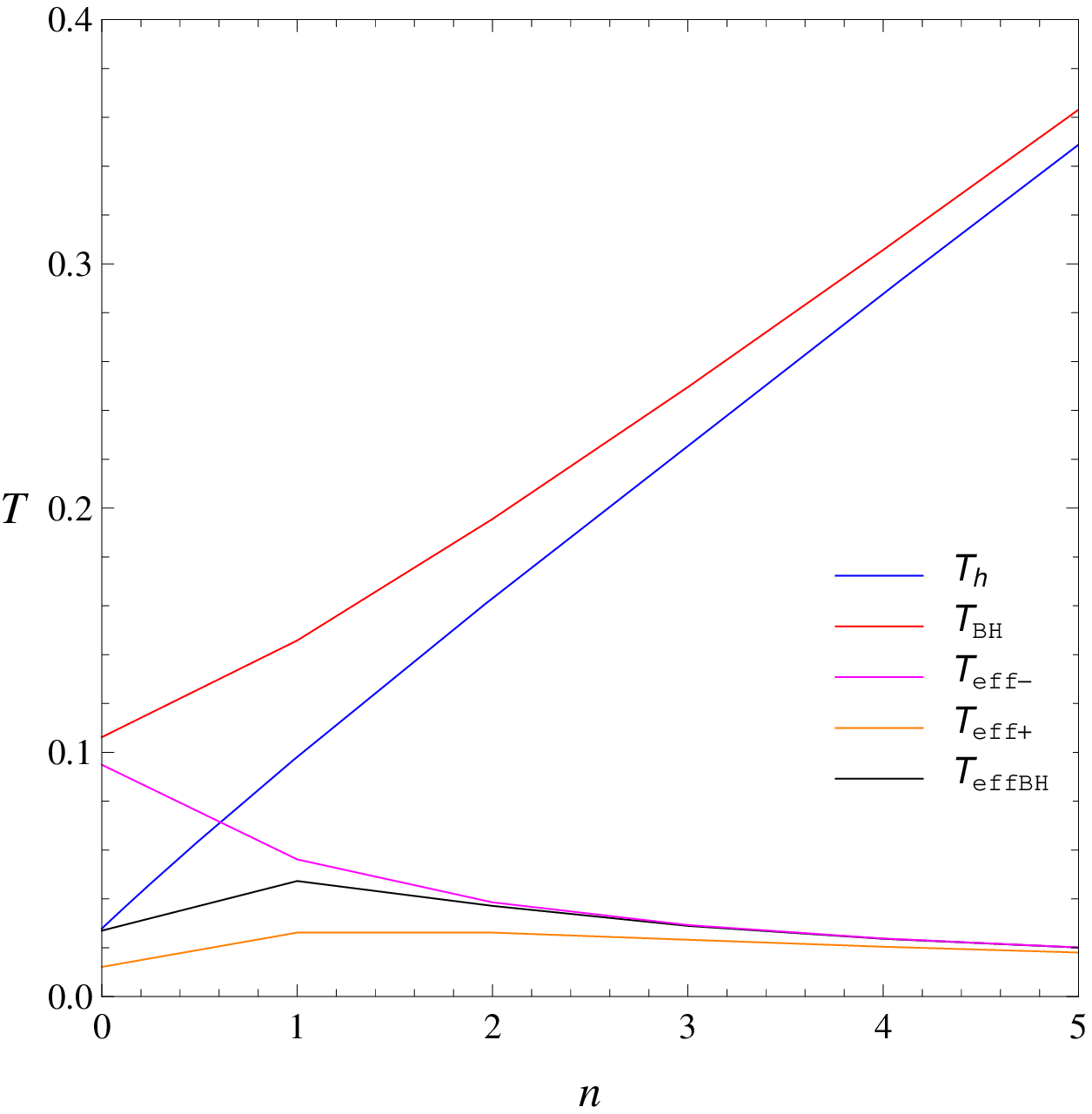}}
\caption{Comparisons of temperatures of higher-dimensional RN-dS black hole with $Q=0.5$, (a) $\Lambda=0.1$, (b) $\Lambda=0.4$.} 
\label{FIGtemp2}
\end{figure}

\begin{figure}[h]

\subfloat[]{\includegraphics[width=0.45\textwidth]{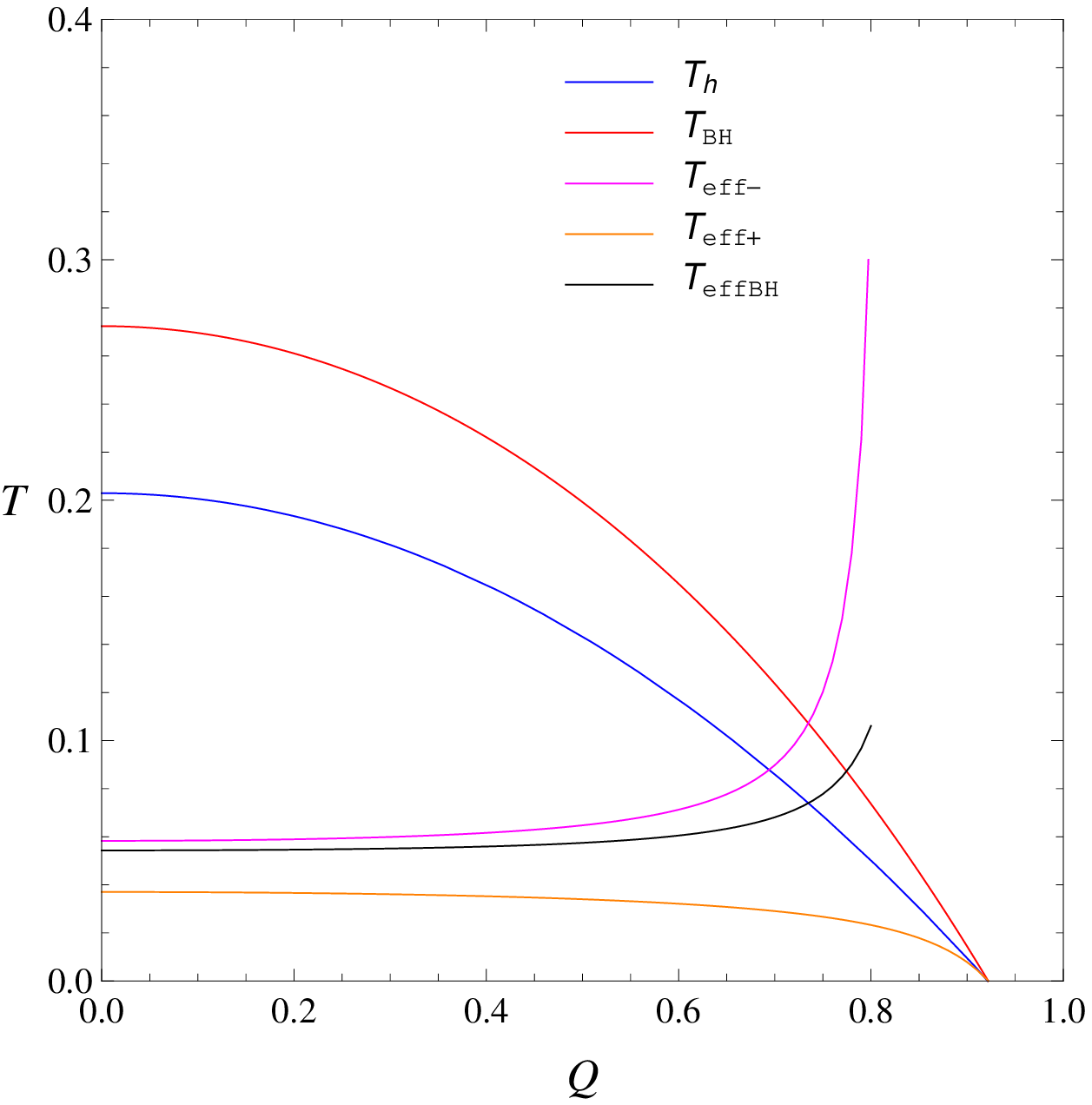}}
\subfloat[]{\includegraphics[width=0.45\textwidth]{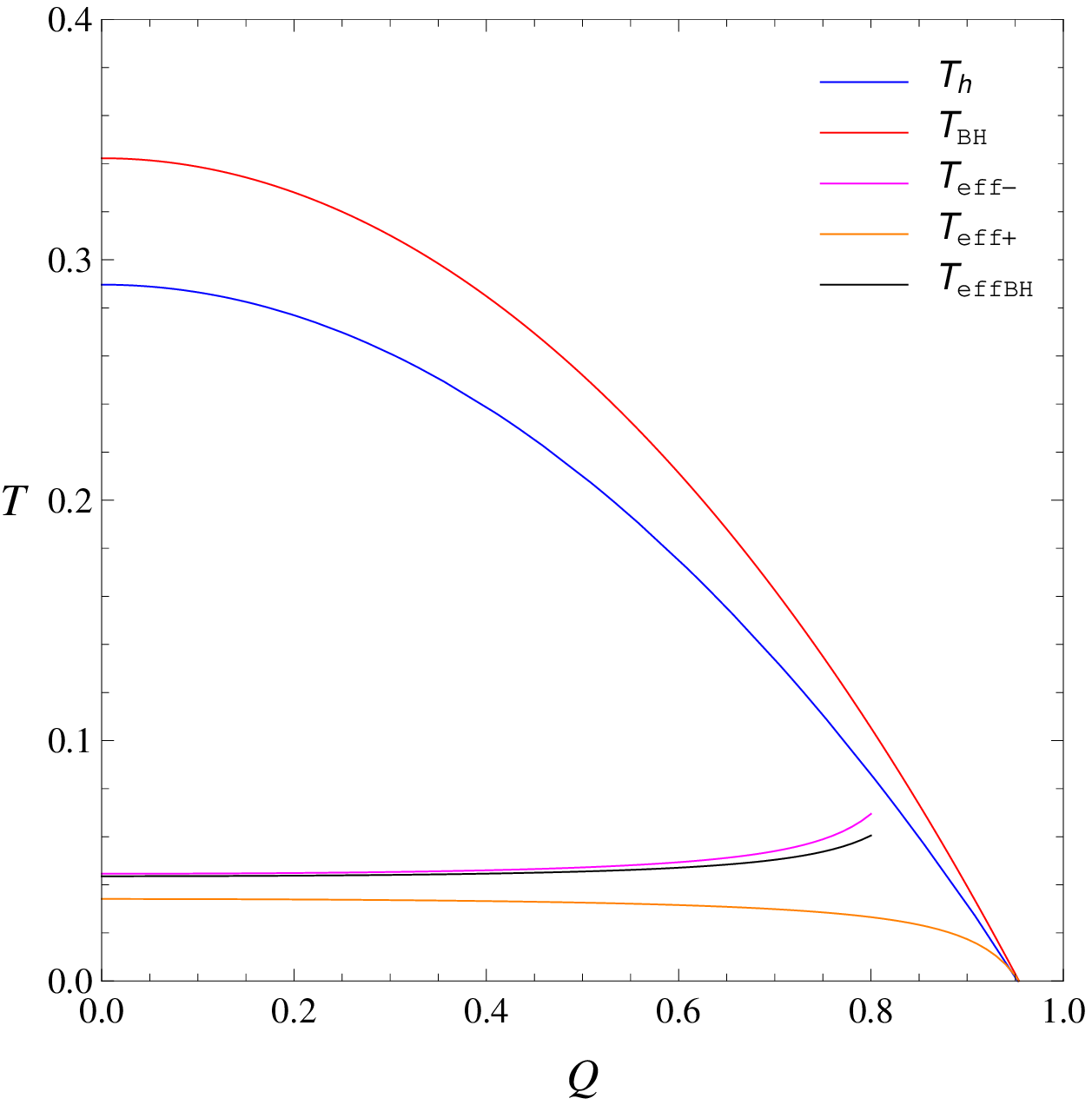}}
\caption{Comparisons of temperatures of higher-dimensional RN-dS black hole with $\Lambda=0.9$, (a) $n=2$, (b) $n=3$.} 
\label{FIGtemp3}
\end{figure}

Now, the effects of cosmological constant $\Lambda$, charge $Q$ on the black hole, and the number $n$ of extra spacetime dimensions on the black hole temperatures $T_h, T_{BH}, T_{eff-}, T_{eff+}, $\\
and $T_{effBH}$ are investigated. These are depicted in Figs.~\ref{FIGtemp1}, \ref{FIGtemp2}, and \ref{FIGtemp3}. Note that in these plots, we chose the background parameters such that $r_0<r_h=1<r_c$.

The temperatures of the black hole are plotted against cosmological constant $\Lambda$ in Fig.~\ref{FIGtemp1}. The conventional temperature $T_h$ decreases monotonically with $\Lambda$ for both small and large value of $n$. When $\Lambda$ approaches the maximum allowed value (reaching the Nairai limit), $T_h$ vanishes. The normalized temperature $T_{BH}$ differs significantly from $T_h$. $T_{BH}$ initially increases with $\Lambda$ before gradually reaching a constant value for low values of $n$. For large values of $n$, $T_{BH}$ continues to increase at lower values of $\Lambda$. As is evident in the figure, $T_h$ and $T_{BH}$ agree only at $\Lambda=0$, as expected. We note that $T_{eff-}$ increases with $\Lambda$. For $n=6$, it is evident that $T_{eff-}$ indicates a sign of divergence closer to the Nairai limit. In contrast, both $T_{eff+}$ and $T_{effBH}$ are zero near the Nairai limit. This is because their numerators converge to zero faster than their denominators. They are similar to $T_{BH}$ in the low-$\Lambda$ regime and resemble $T_h$ in the high-$\Lambda$ regime. 

Fig.~\ref{FIGtemp2} depicts the plot of temperature as a function of the number $n$ of extra dimensions. Varying $n$ does not significantly affect the structure of the horizons ($r_0 < r_h = 1 < r_c$ is always satisfied). Therefore, no infinite jump in any of these temperatures is observed. It is evident that $T_{BH}$ is always larger than $T_h$, as depicted in Fig.~\ref{FIGtemp1}. The difference between these two temperatures is more significant at low values of $n$ and large values of $\Lambda$. All the effective temperatures become nearly identical as $n$ increases. These plots are similar to those presented in \cite{Kanti:2017ubd}.

{
Fig.~\ref{FIGtemp3} depicts the effect of the charge of the black hole on each definition of temperature. With an increase in $Q$, the spacetime structure approaches the extremal limit, $r_0\to r_h$, and $r_h=1$ becomes the smallest real positive root when $Q>Q_{crit}$. Although normalized temperature $T_{BH}$ is significantly larger than $T_h$, they both vanish at a certain critical value $Q_{crit}$. In addition, effective temperature $T_{eff+}$ also becomes zero at $Q=Q_{crit}$  because the numerator of (\ref{Teffp}) becomes zero faster than its denominator owing to $T_h=0$ when $Q$ reaches its critical value. In contrast, $T_{eff-}$ and $T_{effBH}$ exhibit infinite jumps at a certain value of $Q$. This can be understood from the definitions of $T_{eff-}$ (\ref{Teffm}) and $T_{effBH}$ (\ref{TeffBH}). The values of $T_h$ and $T_{BH}$ are generally larger than that of $T_c$. However, as the value of $Q$ increases, the values of $T_h$ and $T_{BH}$ decrease, whereas that of $T_c$ remains unchanged. Therefore, at a certain value of $Q$, temperatures $T_h$ and $T_{BH}$ have the same numerical value as that of $T_c$. Therefore, both $T_{eff-}$ and $T_{effBH}$ diverge. We note that 
$T_{effBH}$ does not exhibit an infinite jump for the higher-dimensional Schwarzschild-dS spacetime \cite{Kanti:2017ubd}. However, an infinite jump in the temperature of the black hole, $T_{effBH}$, is observed when the charge of the black hole is included, as illustrated in these plots.
}


\section{Greybody factor}\label{sect:greybody}

In the preceding section, we explored five different formulae of temperature for higher-dimensional RN-dS black holes in detail. In this section, we derive and investigate the greybody factor of (non-)minimally coupled scalar fields in $(4+n)$-dimensional RN-dS black holes. Particularly, we examine the greybody factor of the brane and bulk scalar fields propagating {on} the RN-dS {background}. Similar studies have been performed for higher-dimensional Schwarzschild-dS \cite{Kanti:2014dxa,Pappas:2016ovo} and four-dimensional RN-dS black holes \cite{Ahmed:2016lou}. For the sake of generality, we calculate the necessary results based on the case of a non-minimally coupled scalar field propagating in bulk spacetimes. 

A non-minimally coupled scalar field in a curved background can be described as 
\begin{align}
\frac{1}{\sqrt{-g}}\partial_{\mu}\left(\sqrt{-g}g^{\mu\nu}\partial_{\nu}\Phi\right) + \epsilon R_n \Phi &= 0, \label{KG}
\end{align}
where $\epsilon$ is a coupling constant and $R_n$ is a higher-dimensional Ricci scalar. 
\begin{align}
R_n &= \frac{2(n+4)}{n+2}\Lambda. \label{RicciHigh}
\end{align}
Considering the factorized spherical symmetric ansatz $\Phi = e^{-i\omega t}R(r)Y(\theta_i,\theta_{n+2})$, where $Y(\theta_i,\theta_{n+2})$ are hyper-spherical harmonics, the scalar field equation (\ref{KG}) can be decomposed into radial and angular parts, as follows
\begin{align}
\frac{1}{r^{n+2}}\frac{d}{dr}\left(fr^{n+2}\frac{dR}{dr}\right) + \left[\frac{\omega^2}{f} - \frac{\ell(\ell+n+1)}{r^2}-\epsilon R_n\right]R &= 0. \label{waveeq}
\end{align}
The angular eigenvalue of the hyperspherical harmonic function is expressed as $\ell(\ell+n+1)$. The $\epsilon$ term can be considered an effective mass term of the scalar field. In the minimally coupled case, the scalar wave equation (\ref{waveeq}) is reduced to the conventional massless Klein-Gordon equation. The projected-on-the-brane scalar field equation can be obtained by setting $n=0$ in the above equation, whereas radial function $f$ continues to be in the same form as (\ref{metricfunction}) {(without setting $n=0$ in $f$)}. In addition, scalar curvature $R_n$ (\ref{RicciHigh}) is replaced by \cite{Kanti:2014dxa}
\begin{align}
R_4 &= \frac{24\Lambda}{(n+2)(n+3)} + \frac{2Mn(n-1)}{r^{n+3}}. \label{RicciHighbrane}
\end{align}

To compute the transmission amplitude of the scalar field or greybody factor, we consider the radial equation in the proximity of the event horizon of the black hole. First, we make the following transformation\cite{Kanti:2014dxa}:
\begin{align}
r \to h(r) &= \frac{f(r)}{1-\bar{\Lambda}r^2}, \nonumber \\
&= 1 - \frac{\left[\left(1-r_h^2\bar{\Lambda}\right) + \frac{Q^2}{(rr_h^2)^{n+1}}(r^{n+1}-r_h^{n+1})\right]}{1-\bar{\Lambda}r^2}\left(\frac{r_h}{r}\right)^{n+1}. \label{transform}
\end{align}
Thus, $h$ varies from $0$, at $r=r_h$, to $1$, when $r\gg r_h$. The following relationship also holds
\begin{align}
\frac{dh}{dr} &\equiv  (1-h)\frac{B(r)}{r\left(1-r^2\bar{\Lambda}\right)},
\end{align}
where $B$ is defined as
\begin{align}
B &= {\scriptstyle \frac{Q^2 \left(\bar{\Lambda} (n+3) r^{n+3}-(n+1) r^{n+1}-2 \bar{\Lambda}  (n+2) r^2 r_h^{n+1}+2 (n+1) r_h^{n+1}\right)-r^{n+1} r_h^{2 n+2} \left(\bar{\Lambda}  r_h^2-1\right) \left(n \left(\bar{\Lambda}  r^2-1\right)+3 \bar{\Lambda}  r^2-1\right)}{Q^2 \left(r_h^{n+1}-r^{n+1}\right)+r^{n+1} r_h^{2 n+2} \left(\bar{\Lambda}  r_h^2-1\right)} }.
\end{align}

Under transformation (\ref{transform}), radial wave equation (\ref{waveeq}) at $r\simeq r_h$ is expressed as follows
\begin{align}
h(1-h)\frac{d^2R}{dh^2} + \left(1-D_h h\right)\frac{dR}{dh} + \left[\frac{\omega^2r_h^2}{B_h^2h} - \frac{\Omega_h\left(1-r_h^2\bar{\Lambda}\right)}{(1-h)B_h^2}\right]R &= 0, \label{nearEHeq}
\end{align}
where 
\begin{align}
\Omega_h &\equiv \ell(\ell+n+1)+\epsilon R^{(h)}_n r_h^2,  \\
B_h &\equiv B(r_h) = (n+1)\left(1-Q^2r_h^{-2(n+1)}\right)-(n+3)r_h^2\bar{\Lambda},  \\
D_h &\equiv 2-\frac{\left(1-r_h^2\bar{\Lambda}\right)}{B_h}\left((n+1) + \frac{r_h B'(r_h)}{B_h}\right).
\end{align}
The scalar curvature $R_n$ was evaluated at $r=r_h$. We redefine the radial field function $R = h^{\alpha}(1-h)^{\beta}H(h)$. Thus, the near-horizon equation (\ref{nearEHeq}) can be substituted into the standard form of a hypergeometric differential equation
\begin{align}
h(1-h)\frac{d^2H}{dh^2} + \left[c-\left(1+a+b\right)h\right]\frac{dH}{dh} - ab H &=0, \label{hyperdiff}
\end{align}
where parameters $a, b$, and $c$ are defined as  
\begin{align}
a &= \alpha+\beta+D_h-1, \\
b &= \alpha+\beta, \\
c &= 1+2\alpha.
\end{align}
Exponents $\alpha$ and $\beta$ are obtained by solving the following equations
\begin{align}
\alpha^2 + \frac{\omega^2 r_h^2}{B_h^2} &= 0, \\
\beta^2 + \beta\left(D_h-2\right) - \frac{\Omega_h\left(1-r^2\bar{\Lambda}\right)}{B_h^2} &= 0.
\end{align}
Thus, the explicit forms of $\alpha$ and $\beta$ are expressed as
\begin{align}
\alpha &=  { \alpha^{(\pm)} \equiv } \pm  \frac{i\omega r_h}{B_h}, \\
\beta &=  { \beta^{(\pm)} \equiv } \frac{B_h(2-D_h) \pm \sqrt{B_h^2\left(2-D_h\right)^2+4\left(1-r_h^2\bar{\Lambda}\right)\Omega_h}}{2B_h}.
\end{align}
The general solution to hypergeometric equation (\ref{hyperdiff}) can be written in terms of hypergeometric function $H$ as
\begin{align}
R_{NH} &= A_1 h^{\alpha}(1-h)^{\beta}H(a,b,c;h) + A_2 h^{-\alpha}(1-h)^{\beta}H(a-c+1,b-c+1,2-c;h), \label{solNBH}
\end{align}
where $A_{1,2}$ are arbitrary constants. At event horizon $h=0$, the general solution reduces to
\begin{align}
R_{NH} &\sim A_1 h^{\alpha} + A_2 h^{-\alpha}.
\end{align}
When $A_{1,2}$ is arbitrary, we have the freedom to choose the sign of $\alpha$.{ With $\alpha=\alpha^{(-)}$, $A_1$ is the coefficient of ingoing wave, whereas the outgoing part is associated with the term with coefficient $A_2$.} Moreover, the convergence of hypergeometric function $H$ requires that Re$(c-a-b)>0$, which implies that $\beta=\beta^{(-)}$. Thereafter, we impose the boundary condition at the event horizon in which only the ingoing wave is allowed. Therefore, we set $A_2=0$, and the near-horizon solution is now
\begin{align}
R_{NH} &\sim A_1 h^{\alpha} = A_1 e^{\frac{-i\omega r_h}{B_h} \ln h}. \label{solNH}
\end{align}
It is possible to obtain an analytic formula of the greybody factor using a matching technique \cite{Kanti:2014dxa,Ahmed:2016lou}. This can be done by considering radial wave equation (\ref{waveeq}) close to the {cosmological horizon}. By considering a region located away from black hole $r_h \ll r_c$, the effects of the mass and charge of the black hole can be neglected. Therefore, the near-{cosmological} horizon equation is significantly simplified. The solution to this equation can be matched with the near-event-horizon solution (\ref{solNH}). However, to fully consider the effect of the cosmological constant and the mass and charge of the black hole, we need to perform a numerical analysis instead of adopting an analytical approach. Our numerical analysis is based on similar analyses performed for higher-dimensional Schwarzschild-dS black holes \cite{Kanti:2014dxa,Kanti:2017ubd}.

We now shift our attention to the near-{cosmological} horizon. Following the same analysis as performed for the near-event-horizon regime, we repeat all the calculations from (\ref{transform}) to (\ref{solNBH})while replacing $r_h, D_h,$ and $B_h$ with $r_c, D_c=D(r_c),$ and $B_c=B(r_c)$, respectively. The solution near the cosmological horizon is expressed as
\begin{align}
R_{NC} &\sim B_1 h^{\tilde{\alpha}} + B_2 h^{-\tilde{\alpha}} = B_1 e^{\frac{-i\omega r_c}{B_c} \ln h} + B_2 e^{\frac{i\omega r_c}{B_c} \ln h}. \label{solNC}
\end{align} 
Integration constants $B_{1,2}$ are defined as amplitudes of the ingoing and outgoing waves at the cosmological horizon. These amplitudes define the transmission probability or greybody factor as \begin{align}
|A|^2 = 1 - \left|\frac{B_2}{B_1} \right|^2. \label{greybodyfactor}
\end{align}
To simplify our numerical calculation, we choose $A_1$ such that $R_{NH}(r_h)=1$ \cite{Kanti:2017ubd}, which plays the role of the boundary condition for the numerical integration. Another boundary condition is determined from the first derivative of $R_{NH}$
\begin{align}
\frac{dR_{BH}}{dr}\rvert_{r=r_h} &\simeq -\frac{i\omega}{f}. \label{BC}
\end{align}
{Radial wave equation (\ref{waveeq}) can now be numerically integrated from the event horizon up to the cosmological horizon using two boundary conditions stated earlier. In practice, we start our numerical integration from very close to the event horizon, that is, $r_h+\delta$, where $\delta$ takes an arbitrary small value. Throughout our computation, $\delta$ was chosen to be $10^{-6}$. In addition, numerical analyses were performed using WOLFRAM'S MATHEMATICA.}

{Integration constants $B_1$ and $B_2$ can be extracted from the near cosmological horizon solution (\ref{solNC}). They are given by}
\begin{align}
B_1 &= \frac{1}{2}\left(\frac{f}{1-r^2\bar{\Lambda}}\right)^{i \omega r_c/B_c}\left[R_{NC} + \frac{i r B_c f}{\omega r_c B \left(1-h\right)}\frac{dR_{NC}}{dr}\right], \\
B_2 &= \frac{1}{2}\left(\frac{f}{1-r^2\bar{\Lambda}}\right)^{- i \omega r_c/B_c}\left[R_{NC} - \frac{i r B_c f}{\omega r_c B \left(1-h\right)}\frac{dR_{NC}}{dr}\right].
\end{align}
{After radial solution $R$ is obtained, the greybody factor can be calculated using (\ref{greybodyfactor}). Note that these coefficients are evaluated at the cosmological horizon. 
}

\subsection{Scalar field on the brane}

\begin{figure}[h]
\subfloat[]{\includegraphics[width=0.45\textwidth]{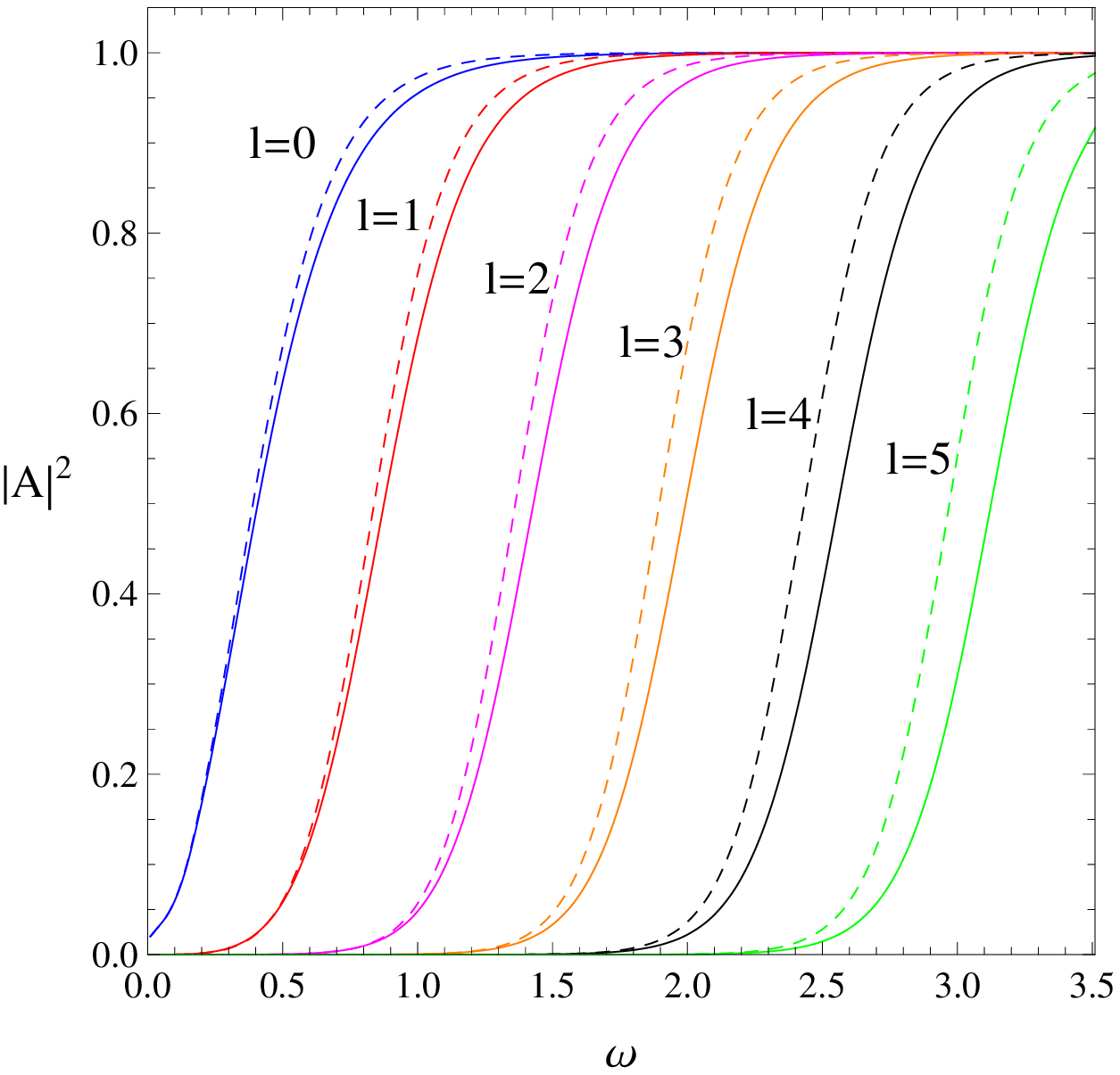}}
\subfloat[]{\includegraphics[width=0.45\textwidth]{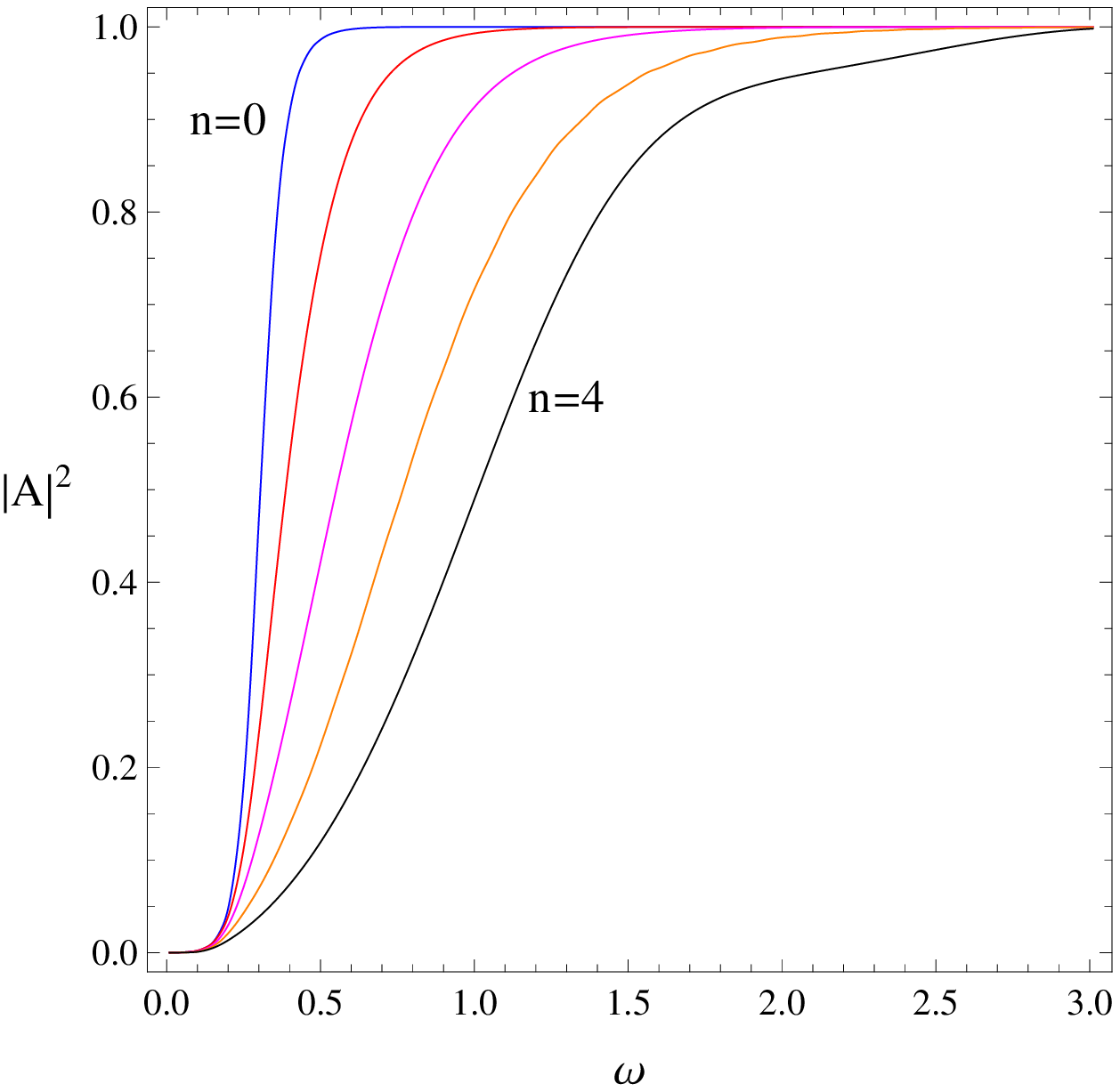}}
\caption{Greybody factor of scalar field on the brane for, (a) $\epsilon=0,n=2,\Lambda=0.05$ with $Q=0$ (solid) and $Q=0.5$ (dashed), (b) $\epsilon=0.5,Q=0.2,\Lambda=0.1$ and $\ell=0$.} 
\label{FIGGreyBrane1}
\end{figure}

First, the greybody factor of the minimally coupled scalar field on the brane, displayed in Fig.~\ref{FIGGreyBrane1}(a), is investigated.
The figure depicts the variation in angular momentum number $\ell$. It is evident that the lowest mode $(\ell=0)$ is the most dominant mode, whereas the higher modes are suppressed when $\ell$ increases. This is expected, because the field mode with spherical symmetry is favored under the {spherical symmetric} spacetime. The effect of the black hole charge $Q$ on the greybody factor is also illustrated in this figure. Charge $Q$ appears to enhance the greybody factor throughout the energy spectrum and for all cases of $\ell$. The effect of the extra spacelike dimension $n$ on the greybody factor of the non-minimally coupled scalar on the brane is depicted in Fig.~\ref{FIGGreyBrane1}(b). In this plot, the greybody factor is plotted for the lowest mode $(\ell=0)$ with coupling parameter $\epsilon=0.3$ and $n=0,1,2,3,4$. It is apparent that the most enhanced greybody factor occurs in four dimensions. The number of extra dimensions clearly suppresses the greybody factor for all ranges of the frequency spectrum. This is also observed in the higher-dimensional Schwarzschild-dS case \cite{Kanti:2014dxa}.

\begin{figure}[h]
\subfloat[]{\includegraphics[width=0.32\textwidth]{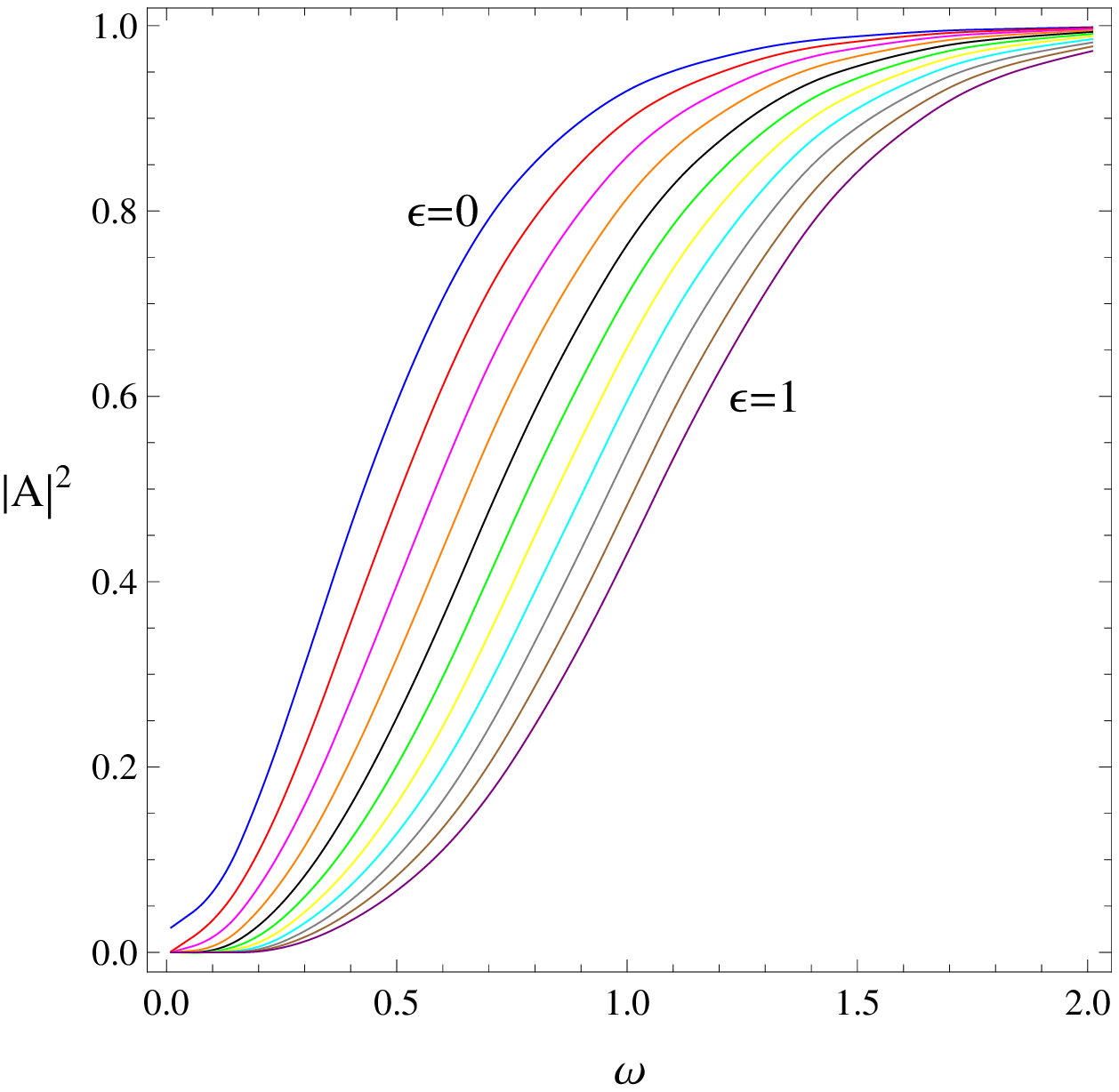} }
\subfloat[]{\includegraphics[width=0.32\textwidth]{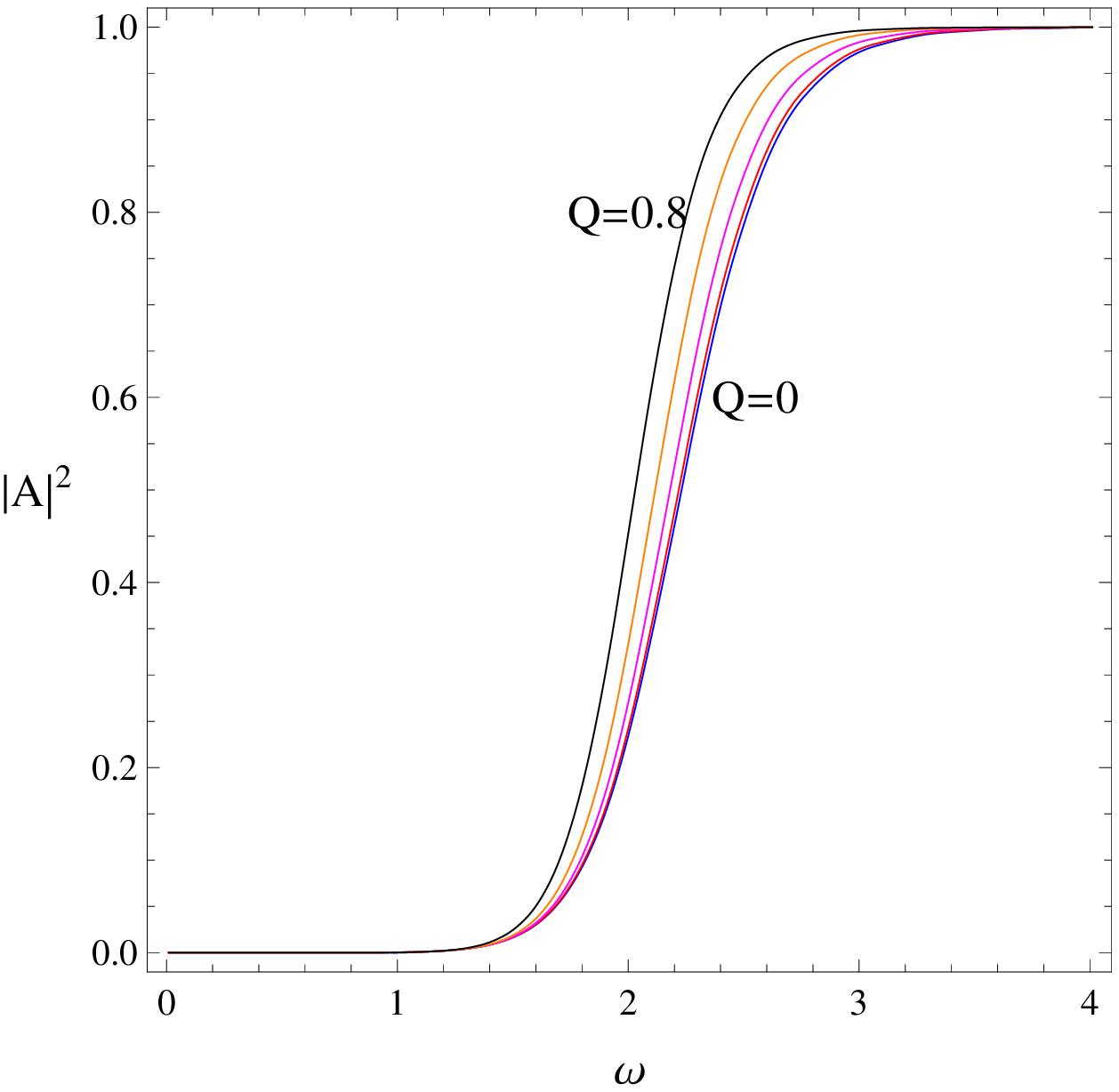} }
\subfloat[]{\includegraphics[width=0.32\textwidth]{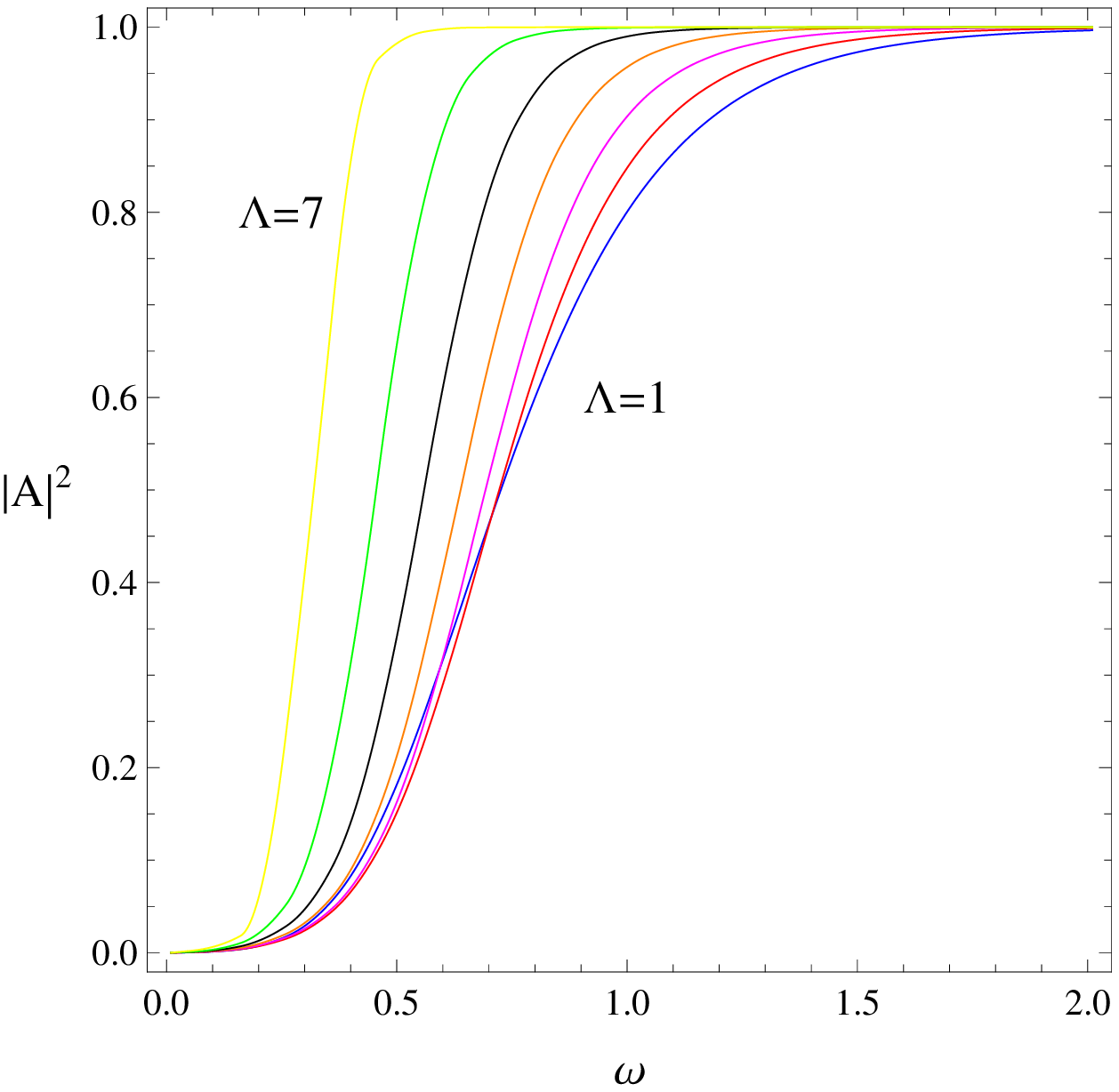} }
\caption{Greybody factor of scalar field on the brane for $n=3$, (a) $Q=0.4,\Lambda=0.1$ and $\ell=0$, (b) $\epsilon=0.4,\Lambda=0.1$ and $\ell=3$, (c) $\epsilon=0.4, Q=0.4$ and $\ell=0$.} 
\label{FIGGreyBrane2}
\end{figure}

Next, we explore the dependence of the greybody factor on various parameters (Fig.~\ref{FIGGreyBrane2}) with the extra dimension $n=3$.
When the coupling constant $\epsilon$ increases, the greybody factor of the lowest partial mode ($\ell=0$) is further suppressed, as depicted in Fig.~\ref{FIGGreyBrane2}(a). This phenomenon was also observed in studies on the greybody factor for a massive scalar field under different conditions \cite{Jung:2004yn,Sampaio:2009ra,Kanti:2010mk}. It can be attributed to the fact that in the equation of motion (\ref{waveeq}), the $R_4$ term plays the same role as the mass term in the scalar field. We also observe that when $\omega$ approaches zero, the greybody factor tends to non-vanishing values in the minimally coupled case, in contrast with the case when $\epsilon \neq 0$, where the transmission amplitude becomes zero. This has also been observed in four-dimensional \cite{Crispino:2013pya} and higher-dimensional Schwarzschild-dS \cite{Kanti:2014dxa,Pappas:2016ovo} setups. The effect of the charge of the black hole on the greybody factor of the scalar field on the brane is illustrated in Fig.~\ref{FIGGreyBrane2}(b). In this figure, the angular momentum number is $\ell=3$ and the coupling parameter is $\epsilon=0.4$ with $Q=0,0.2,0.4,0.6,0.8$. The higher emission mode $\ell$ is chosen to enhance the difference between each curve with a fixed $Q$. When charge $Q$ increases, the transmission amplitude is further suppressed. The greybody factor is significantly different in the intermediate energy spectrum, whereas it apparently agrees in each asymptotic energy spectrum. Finally, we study the response of the greybody factor to the variation in the cosmological constant, as illustrated in Fig.~\ref{FIGGreyBrane2}(c). It is evident that when $\Lambda$ increases, the greybody factor is enhanced. Therefore, from Figs.~\ref{FIGGreyBrane1} and \ref{FIGGreyBrane2}, it is evident that $Q$ and $\Lambda$ increase the scalar transmission amplitude, whereas $\epsilon, n$, and $\ell$ suppress the effect of the greybody factor.

\subsection{Scalar field in bulk}

\begin{figure}[h]
\subfloat[]{ \includegraphics[width=0.45\textwidth]{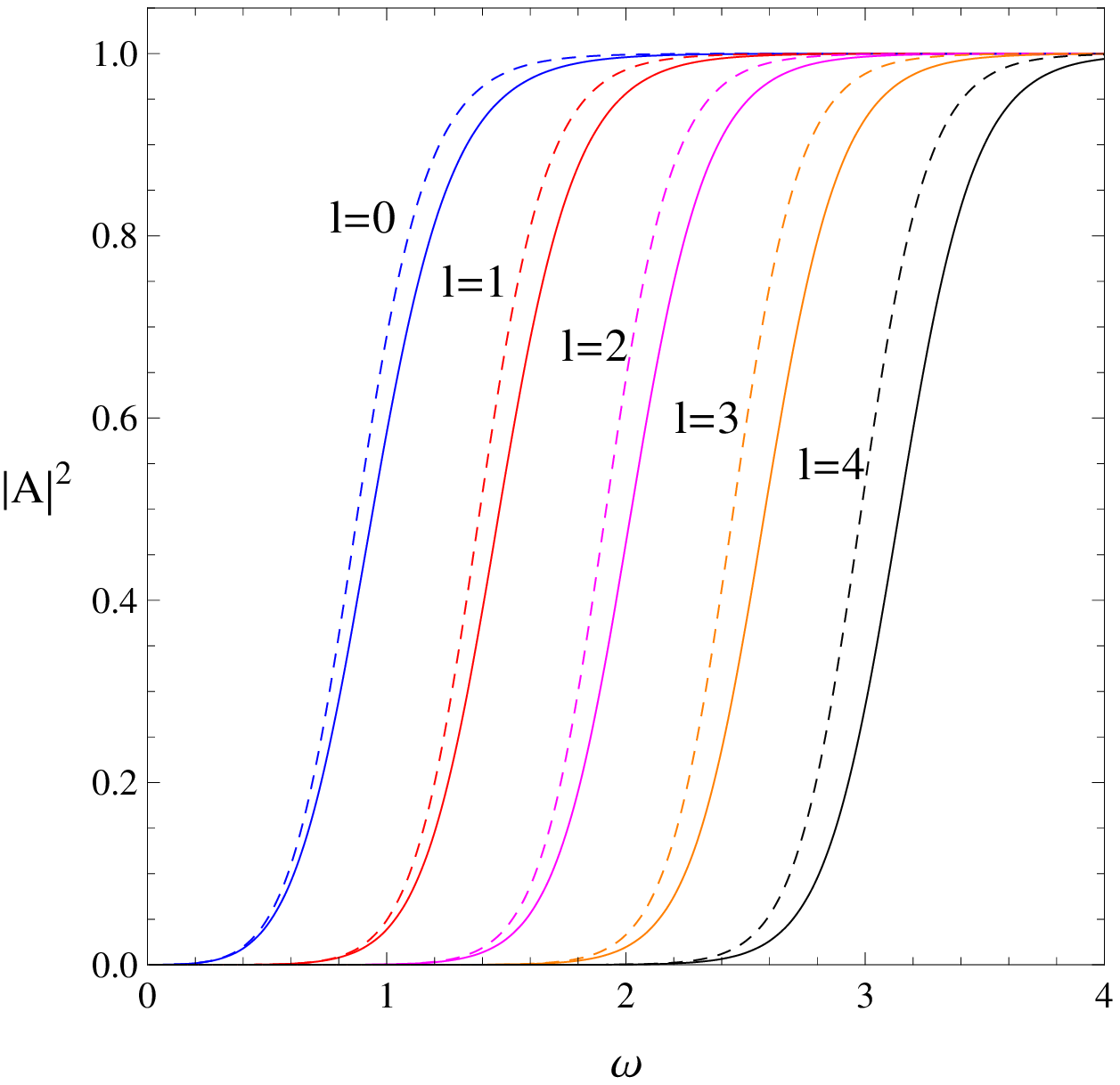} }
\subfloat[]{ \includegraphics[width=0.45\textwidth]{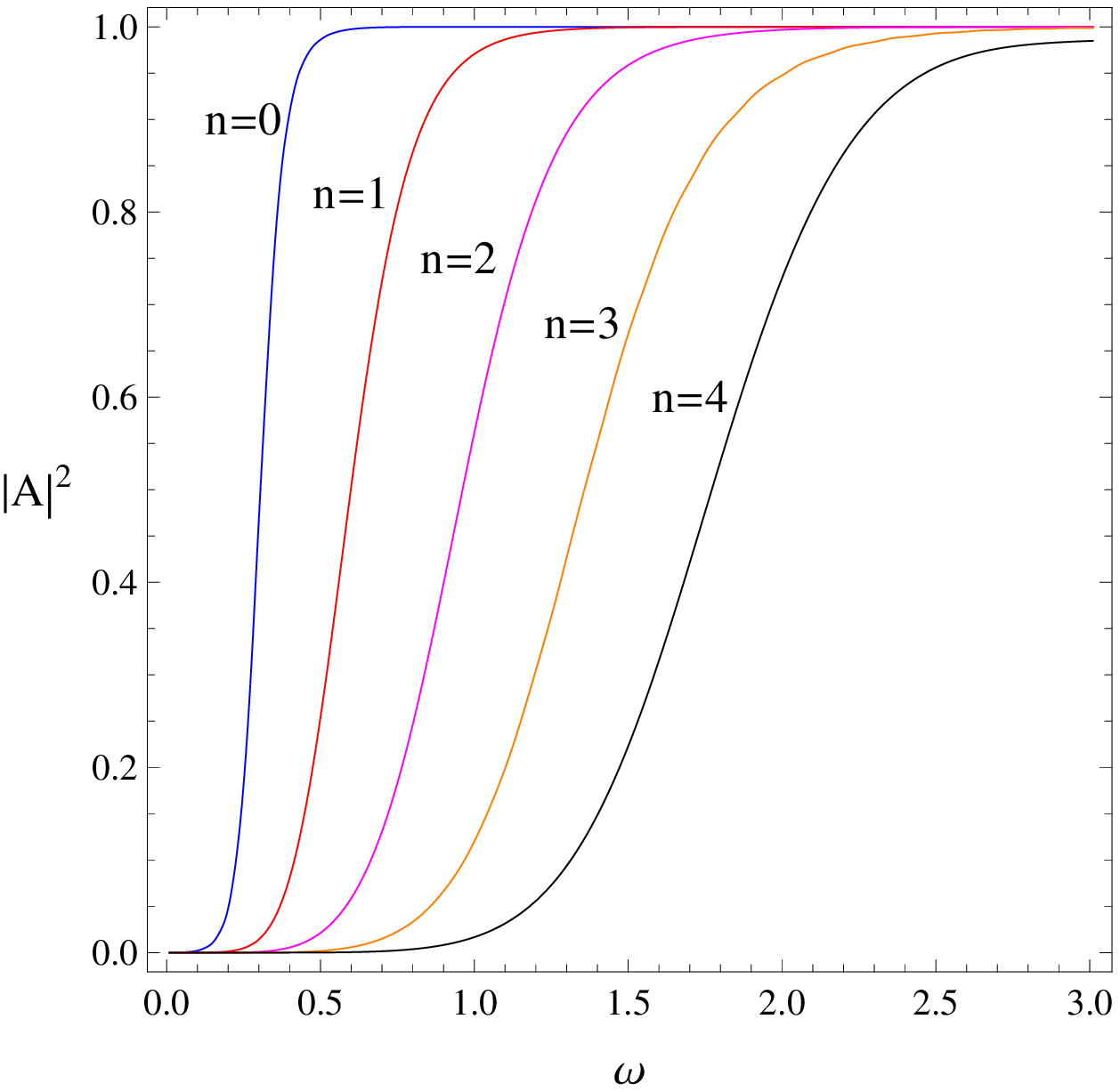} }
\caption{Greybody factor of scalar field in the bulk for, (a) $\epsilon=0,n=2,\Lambda=0.05$ with $Q=0$ (solid) and $Q=0.5$ (dashed), (b) $\epsilon=0.5,Q=0.2,\Lambda=0.1$ and $\ell=0$.} 
\label{FIGGreyBulk1}
\end{figure}

In this subsection, we explore the greybody factor of the bulk scalar field {on} the higher-dimensional RN-dS {spacetime}. For comparison, we chose the same set of parameters as in the brane case. In Fig.~\ref{FIGGreyBulk1}(a), the greybody factor is plotted as a function of the angular quantum number for neutral and charged black holes. Apparently, the increase in $\ell$ suppresses the effect of the greybody factor.
Moreover, the charge of the black hole enhanced the amplitude of the transmission of the scalar field. In addition, this enhancement becomes more significant for higher $\ell$. Fig.~\ref{FIGGreyBulk1}(b) depicts the dependence of the extra spacelike dimension on the greybody factor, which is the most enhanced when the extra dimension vanishes. In higher dimensions, the scalar transmission amplitude is further suppressed. In general, the effects of $\ell$ and $n$ on $|A|^2$ in the bulk are similar to those in the brane case. However, $\ell$ and $n$ appear to have a more suppressed effect on the greybody factor in the bulk case than in the brane case.

\begin{figure}[h]
\subfloat[]{ \includegraphics[width=0.32\textwidth]{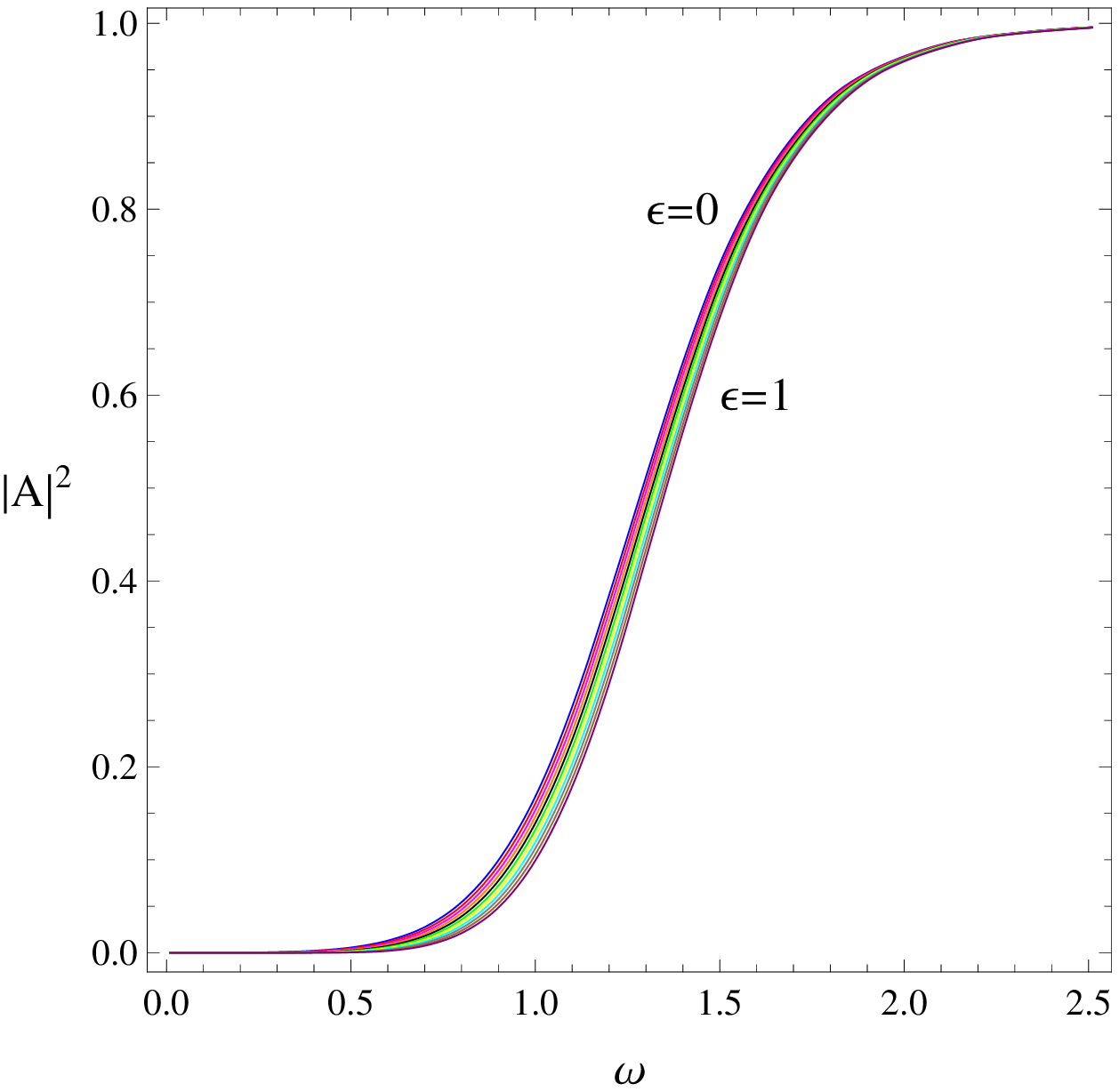} }
\subfloat[]{ \includegraphics[width=0.32\textwidth]{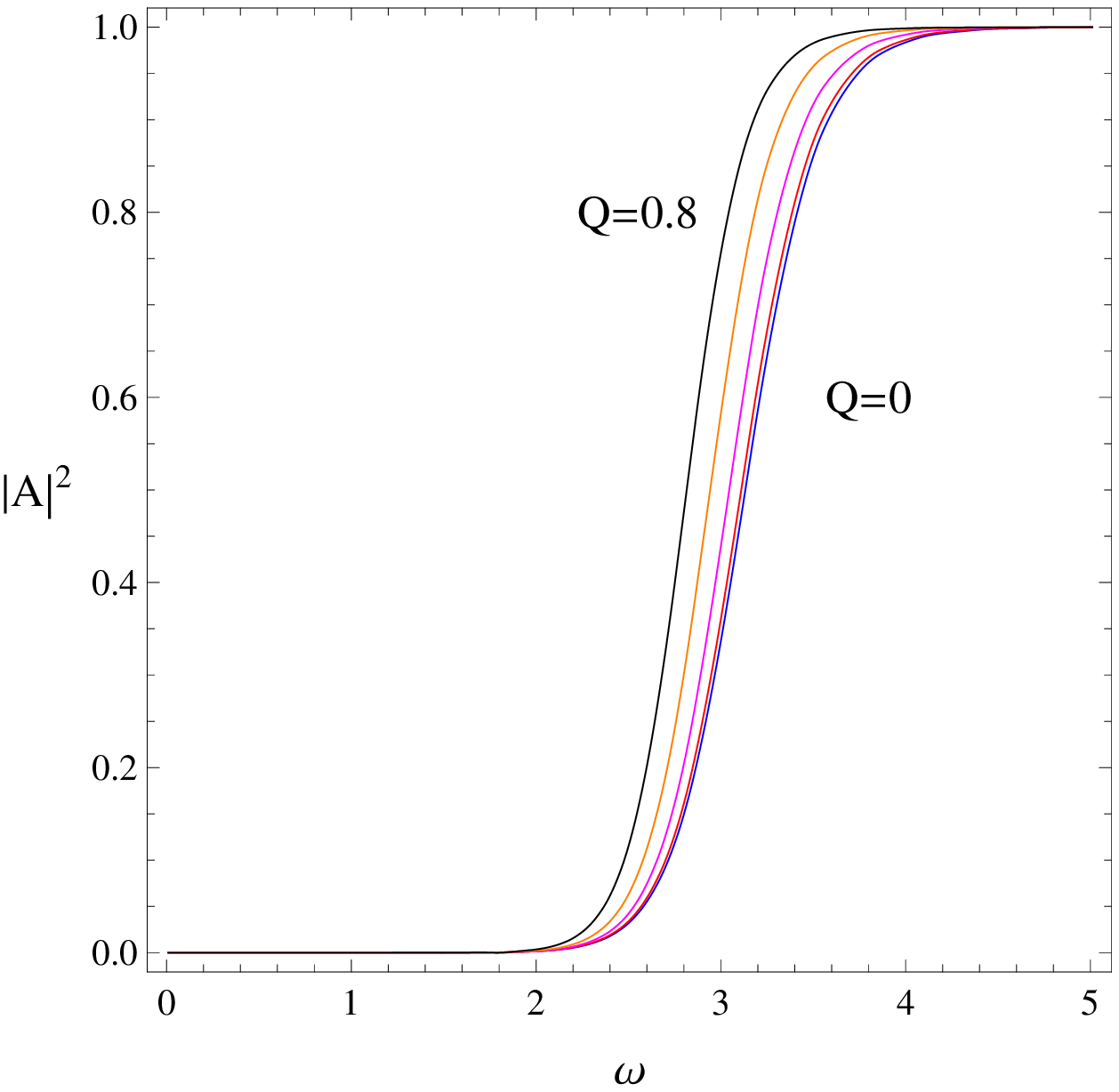} }
\subfloat[]{ \includegraphics[width=0.32\textwidth]{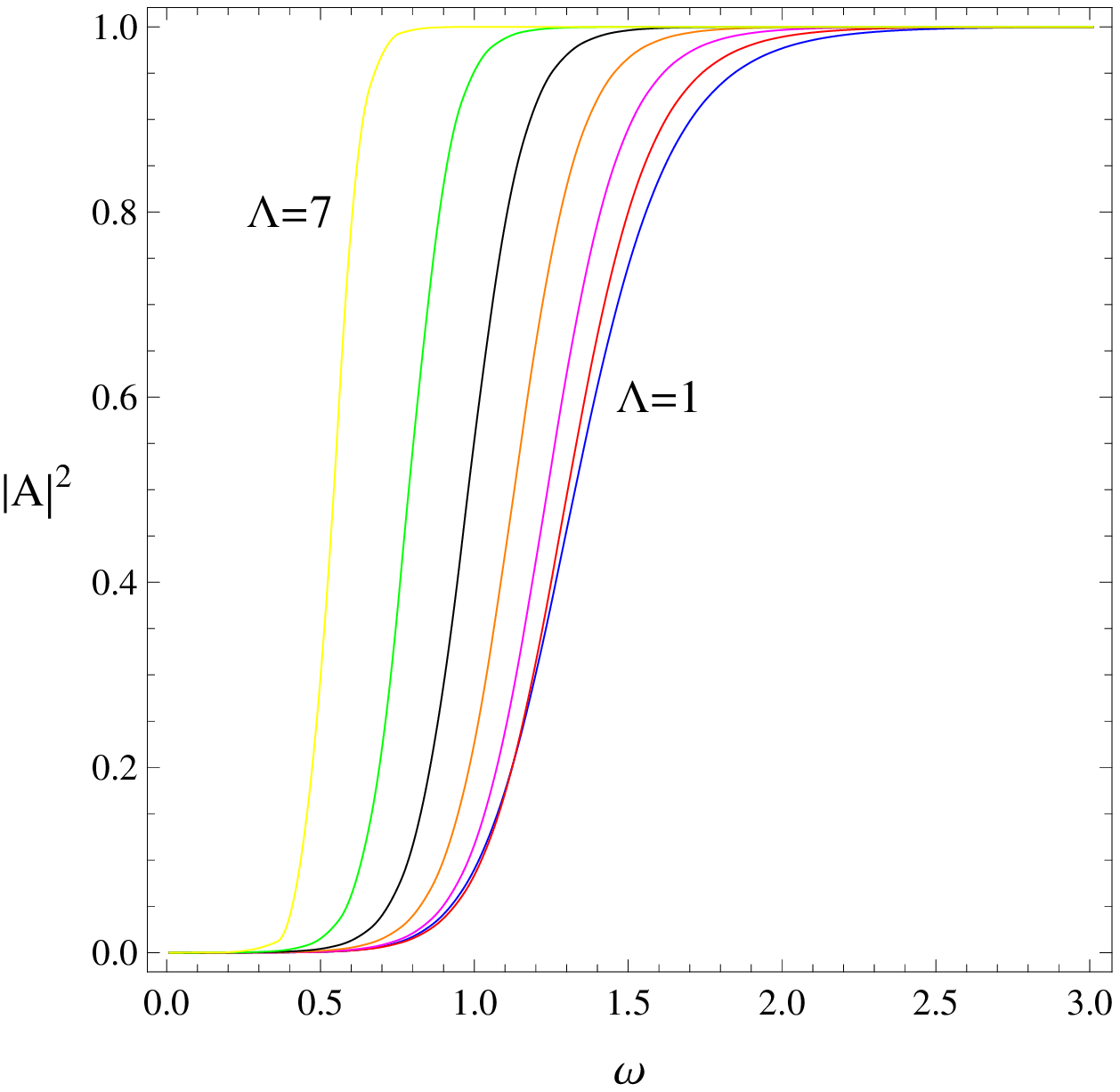} }
\caption{Greybody factor of scalar field in the bulk for $n=3$, (a) $Q=0.4,\Lambda=0.1$ and $\ell=0$, (b) $\epsilon=0.4,\Lambda=0.1$ and $\ell=3$, (c) $\epsilon=0.4, Q=0.4$ and $\ell=0$.} 
\label{FIGGreyBulk2}
\end{figure}

Fig.~\ref{FIGGreyBulk2} depicts the effects of parameters $\epsilon, Q$, and $\Lambda$, with the extra dimension $n=3$. The coupling constant to the Ricci scalar plays the same role as the effective scalar field mass.
Therefore, we expect that the greybody factor will be further suppressed when the coupling parameter increases, similar to the brane case. In the bulk scenario, we also observe the variation depicted in Fig.~\ref{FIGGreyBulk2}(a). However, the suppression effect of the coupling constant is less significant than that of the brane case. The dependences of the charge of the black hole and cosmological constant on $|A|^2$ are generally identical to those in the brane case, that is, both enhance the greybody factor; see Figs.~\ref{FIGGreyBulk2}(b) and \ref{FIGGreyBulk2}(c). In comparison with the brane scenario, we observe that the greybody factor is further suppressed in the low-energy limit. In general, we see the same variation as in the brane case, that is, $Q$ and $\Lambda$ improve the greybody factor, whereas $\epsilon, n$, and $\ell$ suppress the effect of $|A|^2$.

\section{Energy emission rates of scalar field}\label{sect:EER}

In this section, we discuss the differential energy emission rates of the scalar field {propagating in} the higher-dimensional RN-dS {spacetime}. Particularly, we investigate the emission of (non-)minimally coupled scalar both on the brane and in the bulk. We also study the dependences of the charge of the black hole, cosmological constant, and coupling parameter on the energy emission rates on the brane and bulk.

The differential energy emission rate of the scalar field is defined as \cite{Harris:2003eg,Kanti:2004nr,Kanti:2005ja}
\begin{align}
\frac{d^2E}{dt d\omega} &= \frac{1}{2\pi}\sum_{\ell}\frac{N_{\ell}|A|^2\omega}{\text{exp}\left(\omega/T\right)-1}, \label{emittbrane}
\end{align}
where $\omega$ is the energy of the emitted particle and $T$ is the temperature of the black hole. The multiplicity of states is expressed as \cite{Kanti:2005ja}
\begin{align}
N_{\ell} = \left\{ \begin{array}{lr}
 \left(2\ell+1\right)  \hspace{3.1cm} \mbox{ brane}, \\
 \frac{\left(2\ell+n+1\right)\left(\ell+n\right)!}{\ell!\left(n+1\right)!} \hspace{2.4cm} \mbox{ bulk}.
       \end{array} \right.
\end{align}
Formula (\ref{emittbrane}) can be determined when the greybody factor $|A|^2$ and temperature of the black hole are known. The greybody factor is calculated using the numerical routine discussed in the preceding section. The temperature of the black hole can be either $T_h,T_{BH},T_{eff-},T_{eff+}$,or $T_{effBH}$, as discussed earlier. Furthermore, we compare the energy emission rates for each definition of the temperature of the black hole.

\subsection{Energy emission on brane}

\begin{figure}[h]
\subfloat[]{ \includegraphics[width=0.45\textwidth]{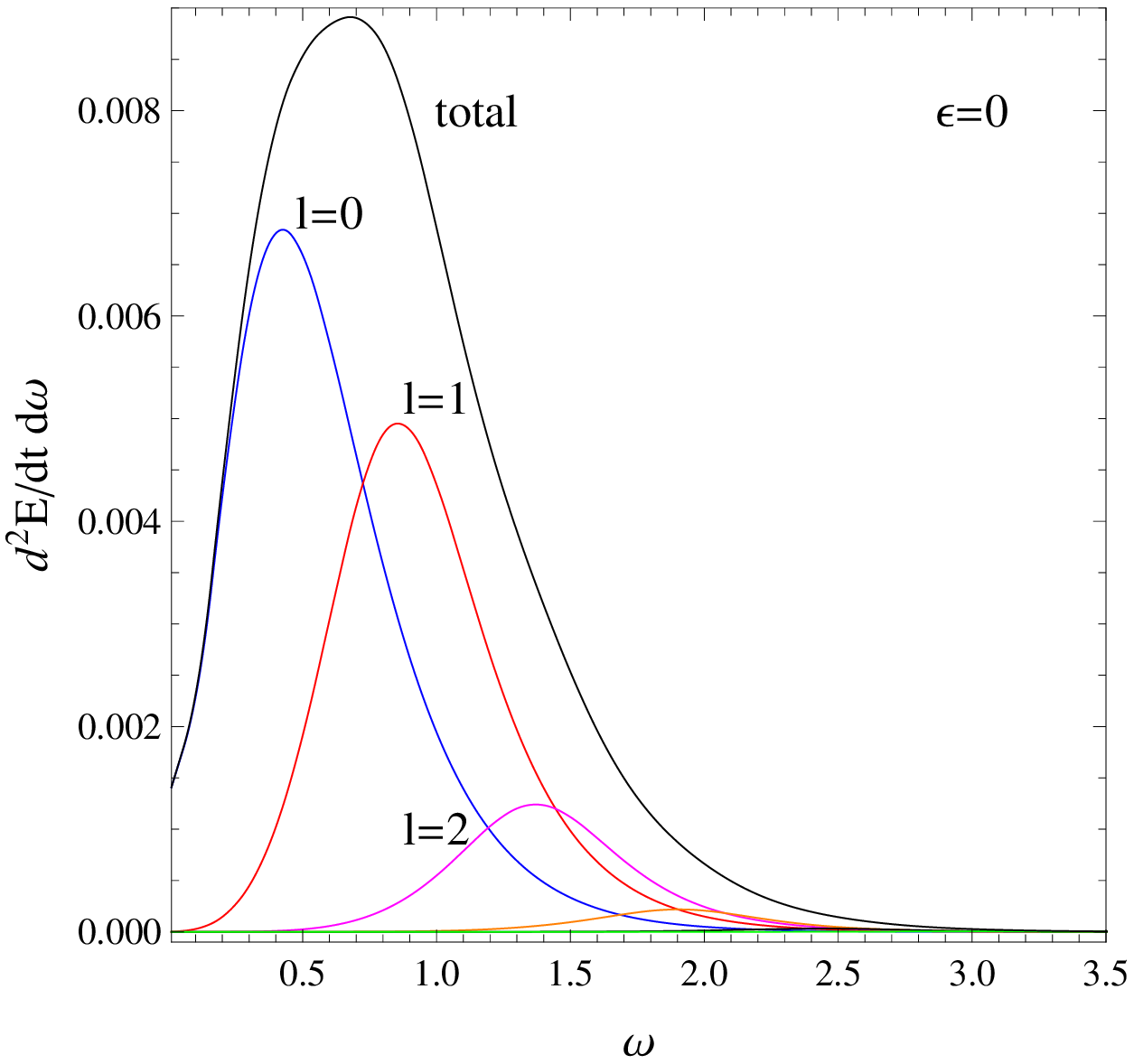} }
\subfloat[]{ \includegraphics[width=0.45\textwidth]{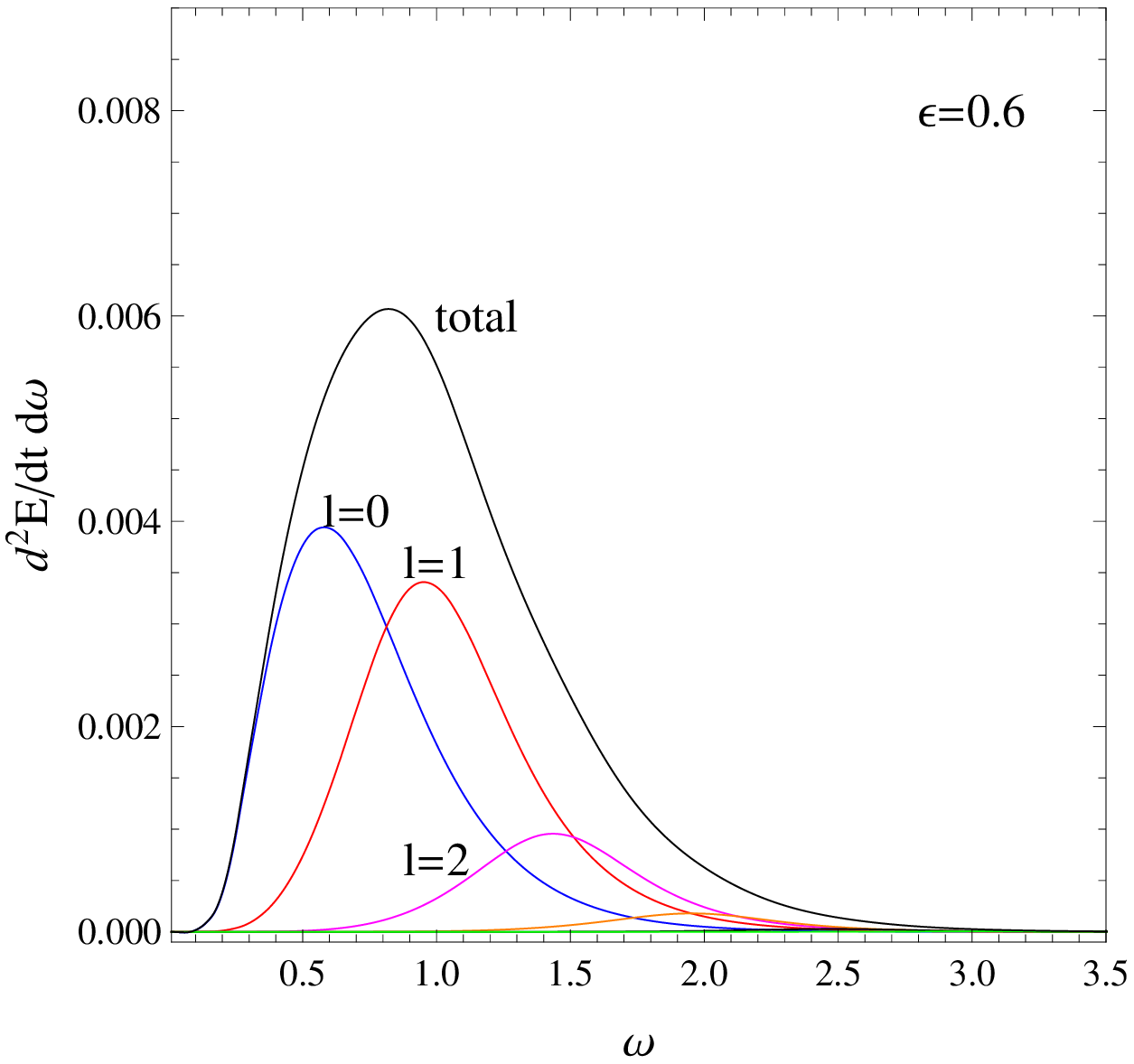} }
\caption{Energy emission rate for scalar field on the brane for $n=2,Q=0.3,\Lambda=0.1$, (a) $\epsilon=0$, (b) $\epsilon=0.6$, with the first five modes $\ell=0,1,2,3,4,5$.} 
\label{FIGEERbrane1}
\end{figure}

We start with the energy emission of the scalar field on the brane. In Fig.~\ref{FIGEERbrane1}, we depict the contribution of the dominant modes of the scalar field to the total energy emission. Note that the Bousso-Hawking temperature (\ref{TBH}) is chosen. The total emission rate was computed for modes $\ell=0-5$.
We observe that the lowest mode ($\ell=0$) is the most dominant mode in the total energy emission, whereas the higher modes contribute less significantly. Therefore, we ignore the $\ell>5$ modes in the calculation of (\ref{emittbrane}) in this study, unless stated otherwise. We also observe the suppression of the energy emission rate as the coupling constant increases. This is similar to what we found for the greybody factor discussed earlier. It is apparent that the emission rate of the lower modes is more affected by the suppression of $\epsilon$ than that of the higher modes. Finally, in the limit $\omega\to 0$, the non-vanishing value of the lowest mode $\ell=0$ is inherited from the fact that the greybody factor is nonzero as the frequency approaches zero for the minimally coupled scalar field.

\begin{figure}[h]
\subfloat[]{ \includegraphics[width=0.45\textwidth]{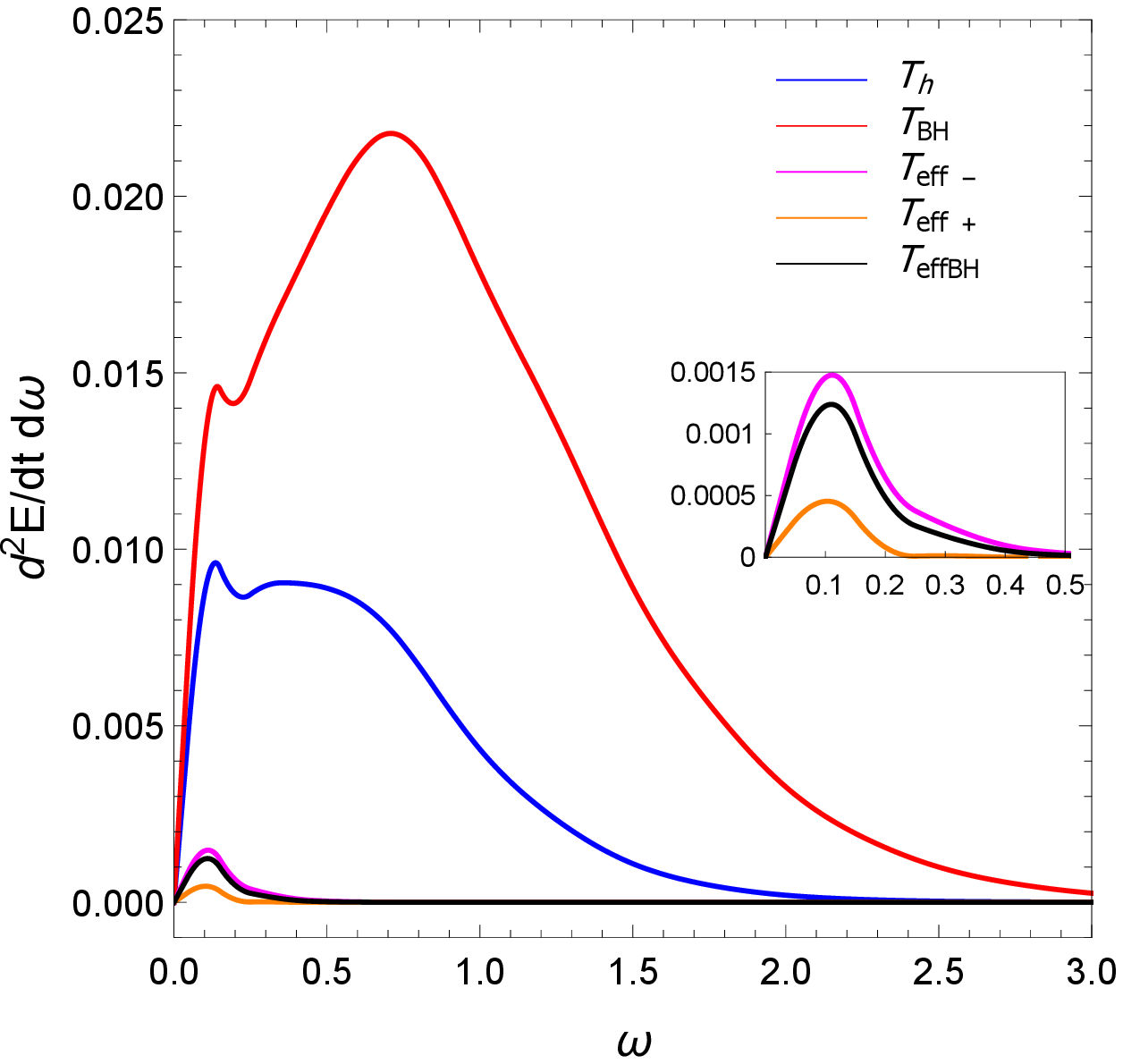} }
\subfloat[]{ \includegraphics[width=0.45\textwidth]{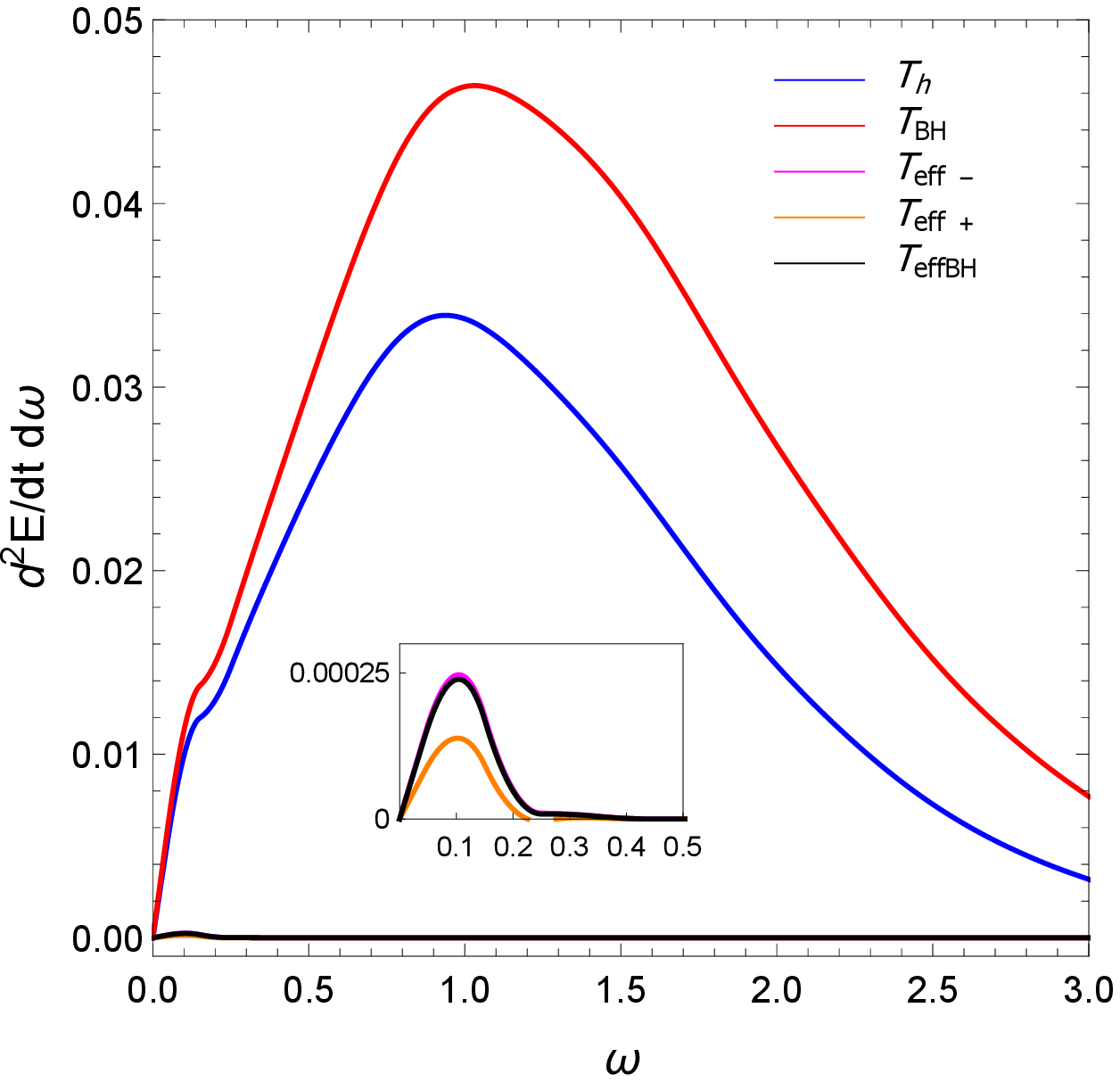} }
\caption{Comparisons of energy emission rate for scalar field on the brane for each temperatures with $\epsilon=0.01,Q=0.1,\Lambda=1$ (a) $n=2$, (b) $n=4$.} 
\label{FIGEERbrane2}
\end{figure}

Next, the comparison of the energy emission rates of the non-minimally coupled scalar field for each black hole temperature is made. The differential energy emission rates for higher-dimensional RN-dS black holes with charge $Q=0.1$ and cosmological constant $\Lambda=1$ are displayed in Fig.~\ref{FIGEERbrane2}. For a six-dimensional black hole, the energy emission rates vanish as $\omega \to 0$. When the frequency increases, the emission rates reach their peaks at a certain value of $\omega$ before decreasing to zero at a larger value of $\omega$. This bell-shaped curve of the energy emission rate is typical and is found in similar studies on the Schwarzschild-dS black holes \cite{Kanti:2017ubd,Crispino:2013pya}. The emission rates of energy for $T_h$ and $T_{BH}$ are relatively higher than those for the effective definitions of the temperatures of the black holes, $T_{eff-},T_{eff+}$, and $T_{effBH}$. This is because $T_h$ and $T_{BH}$ are generally higher than $T_{eff-},T_{eff+}$, and $T_{effBH}$, thus causing a smaller denominator in (\ref{emittbrane}). The subplot depicts similar variations in the curves of $T_{eff-}$ and $T_{effBH}$. 

For an eight-dimensional black hole (Fig.~\ref{FIGEERbrane2}(b)), differential energy emission rates are enhanced for the $T_h$ and $T_{BH}$ curves. Whereas the greybody factor is generally suppressed, $T_h$ and $T_{BH}$ increase with $n$, as depicted in Fig.~\ref{FIGtemp2}. The increasing temperature dominates the suppression of the greybody factor, thus enhancing these curves. Conversely, for effective temperatures, the energy emission rates decrease when the number of extra dimensions increases because $T_{eff-},T_{eff+}$,and $T_{effBH}$ decrease monotonically with $n$. Therefore, the denominator of (\ref{emittbrane}) increases as $n$ decreases. Moreover, the $T_{eff-}$ and $T_{effBH}$ curves become more identical because both $T_{eff-}$ and $T_{effBH}$ share a common feature when $n$ increases.

\begin{figure}[h]

\subfloat[]{ \includegraphics[width=0.45\textwidth]{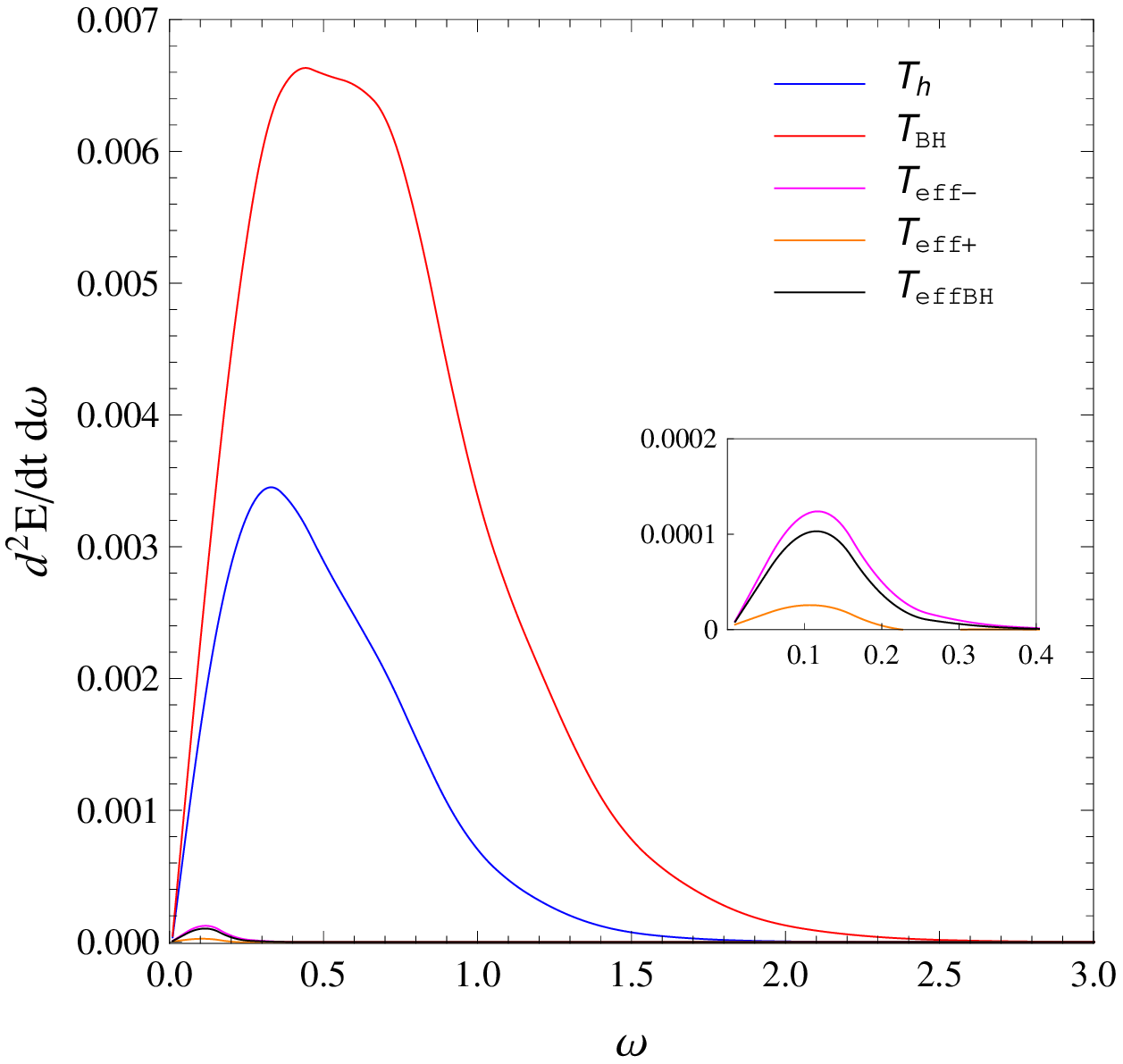} }
\subfloat[]{ \includegraphics[width=0.45\textwidth]{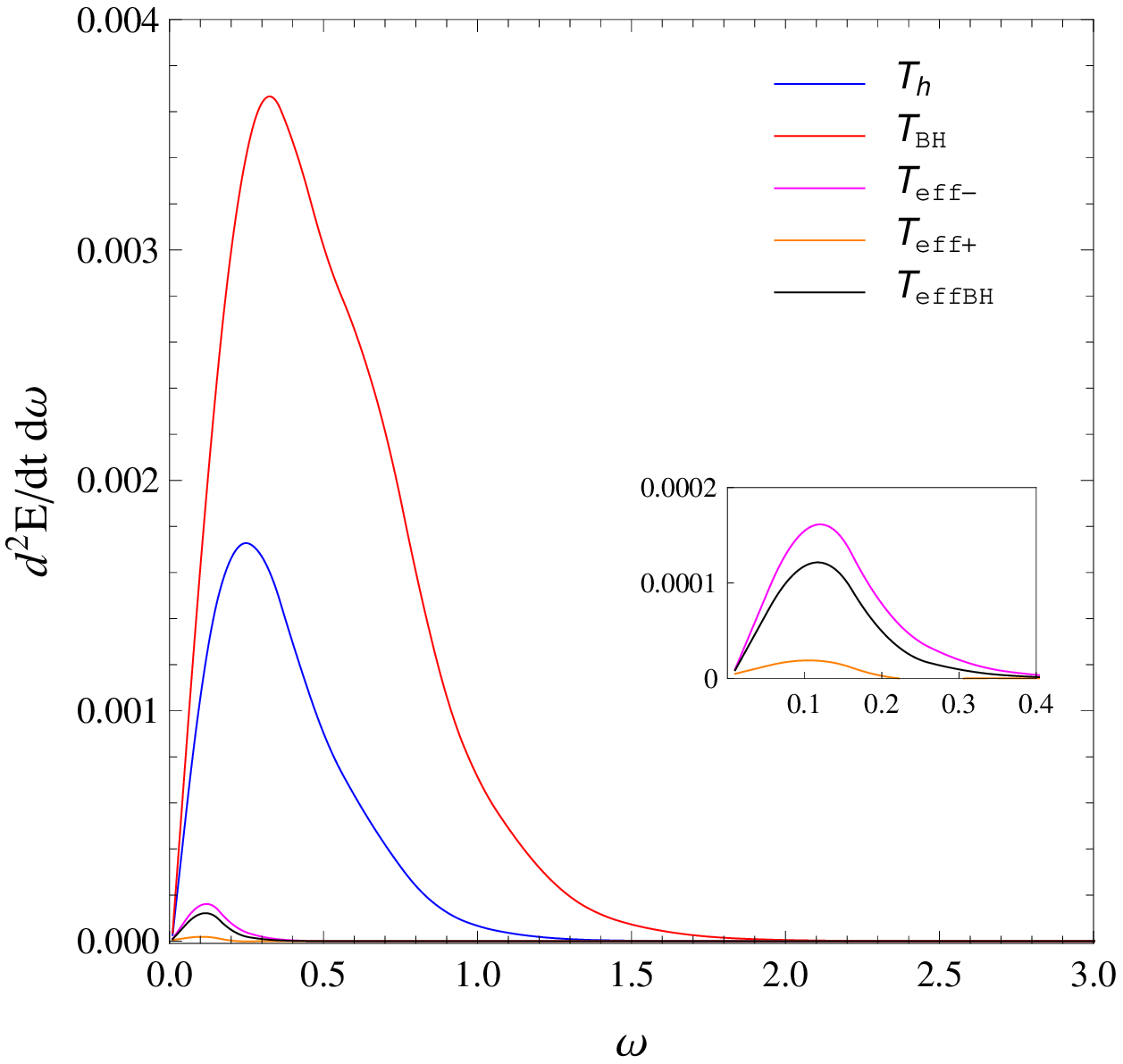} }
\caption{Comparisons of energy emission rate for scalar field on the brane for each temperatures with $\epsilon=0.1,n=1,\Lambda=0.25$, (a) $Q=0.1$, (b) $Q=0.5$.} 
\label{FIGEERbrane3}
\end{figure}

Fig.~\ref{FIGEERbrane3} depicts the energy emission rates of the non-minimally coupled scalar field on the five-dimensional RN-dS black hole. First, we notice that the emission rate vanishes in the low-energy regime. The emission curves also exhibit the typical bell shape for the five different definitions of the temperature of the black hole. In contrast with the previous plot, the increasing charge $Q$ suppresses the $T_h, T_{BH}$, and $T_{eff+}$ energy emission curves. However, it enhances the $T_{eff-}$ and $T_{effBH}$ curves instead. When the charge of the black hole increases, the greybody factor is typically enhanced. However, $T_h, T_{BH}$, and $T_{eff+}$ decrease with respect to the charge of the black hole, as depicted in Fig.~\ref{FIGtemp3}, whereas $T_{eff-}$ and $T_{effBH}$ increase. These affect the denominator of (\ref{emittbrane}) such that it becomes larger (for $T_h, T_{BH}$, and $T_{eff+}$) and smaller (for $T_{eff-}$ and $T_{effBH}$). Notably, the power spectrum of brane emission becomes significantly narrower as the charge of the black hole increases.

\begin{figure}[h]

\subfloat[]{ \includegraphics[width=0.45\textwidth]{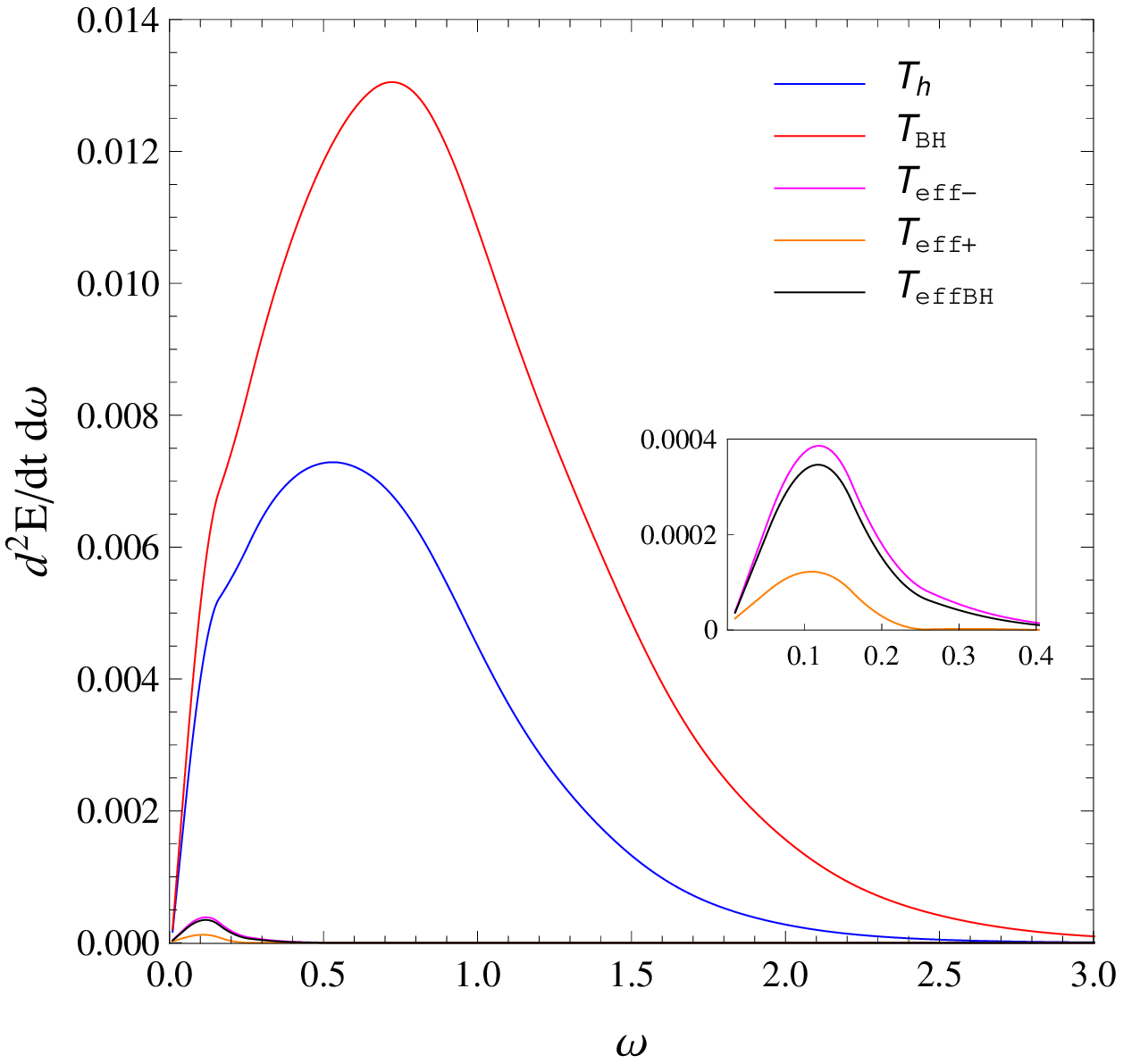} }
\subfloat[]{ \includegraphics[width=0.45\textwidth]{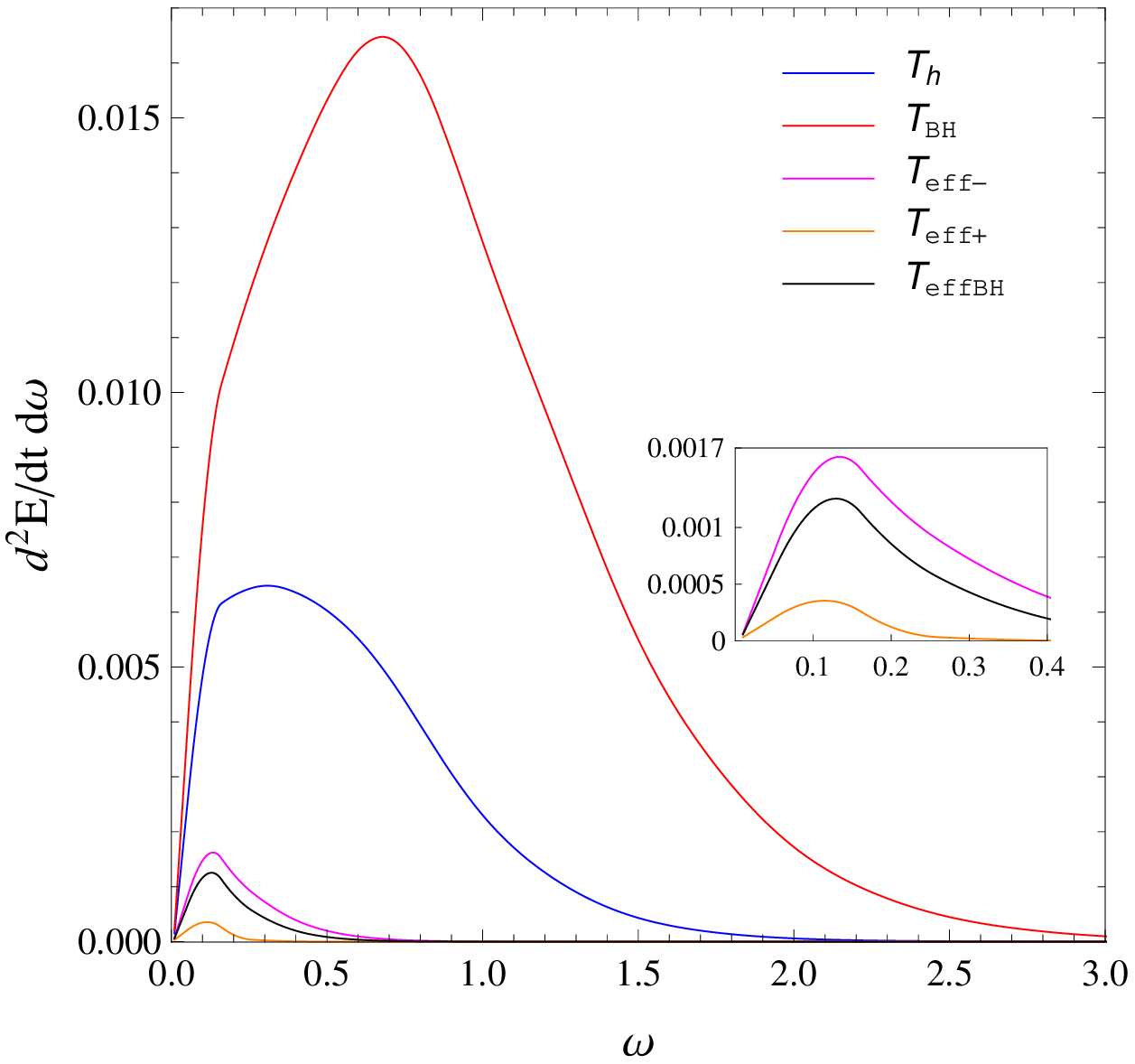} }
\caption{Comparisons of energy emission rate for scalar field on the brane for each temperatures with $\epsilon=0.05,n=3,Q=0.5$, (a) $\Lambda=1$, (b) $\Lambda=2$.} 
\label{FIGEERbrane4}
\end{figure}

Fig.~\ref{FIGEERbrane4} illustrates the effect of the cosmological constant on the energy emission rates for seven-dimensional RN-dS black holes. Similar to the previous case, the bell-shaped curves and vanishing initial values are also seen. This figure indicates that when the cosmological horizon decreases, all the curves of differential energy emission rates, except the traditional black hole temperature curve $T_h$, are enhanced. From Fig.~\ref{FIGtemp1}, we deduce that the $T_h$ line is the only one that decreases monotonically with $\Lambda$, whereas the others increase in the intermediate-$\Lambda$ regime. This explains why suppression appears only in the $T_h$ curve.

\subsection{Energy emission in bulk}

\begin{figure}[h]
\subfloat[]{ \includegraphics[width=0.45\textwidth]{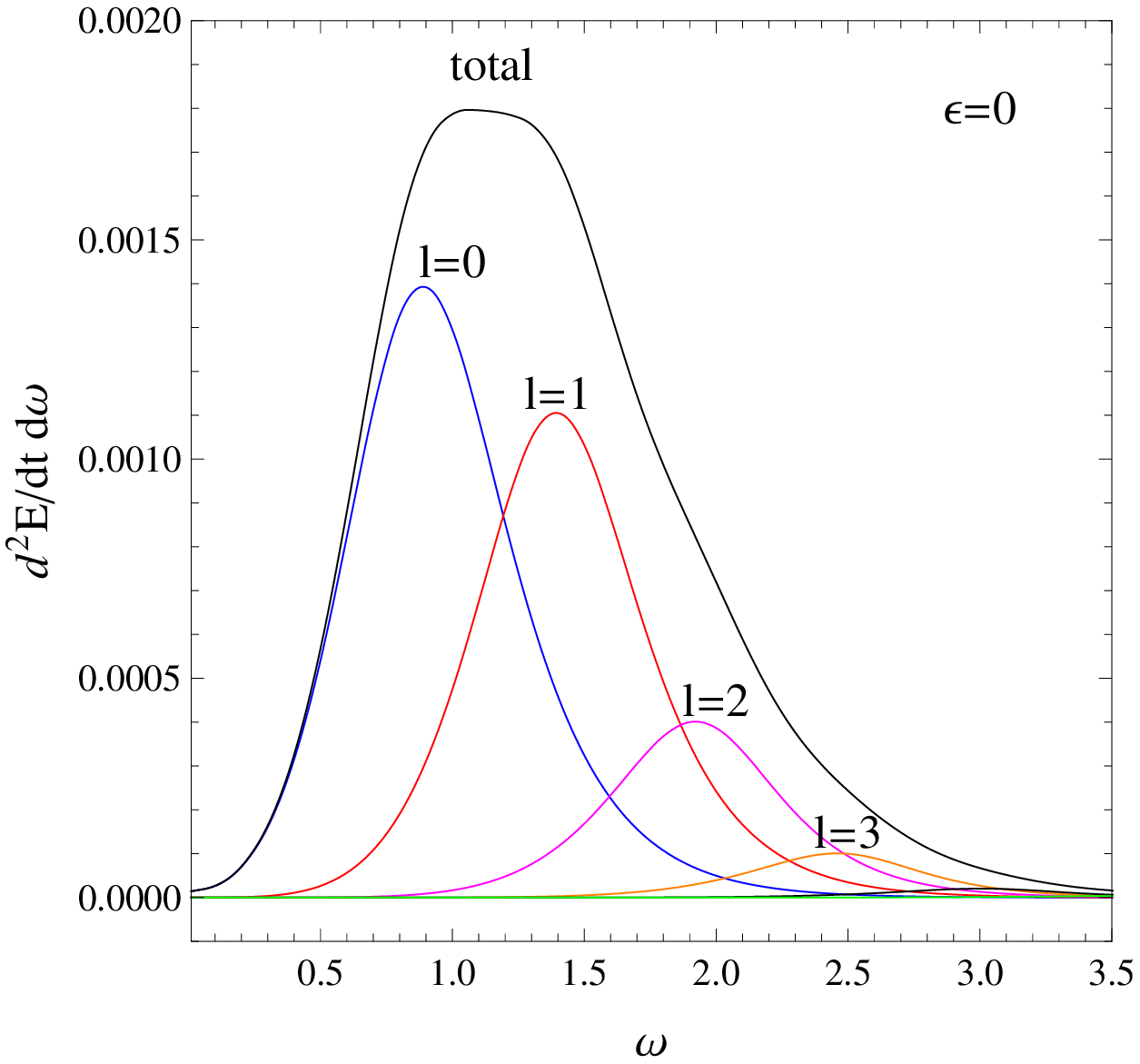} }
\subfloat[]{ \includegraphics[width=0.45\textwidth]{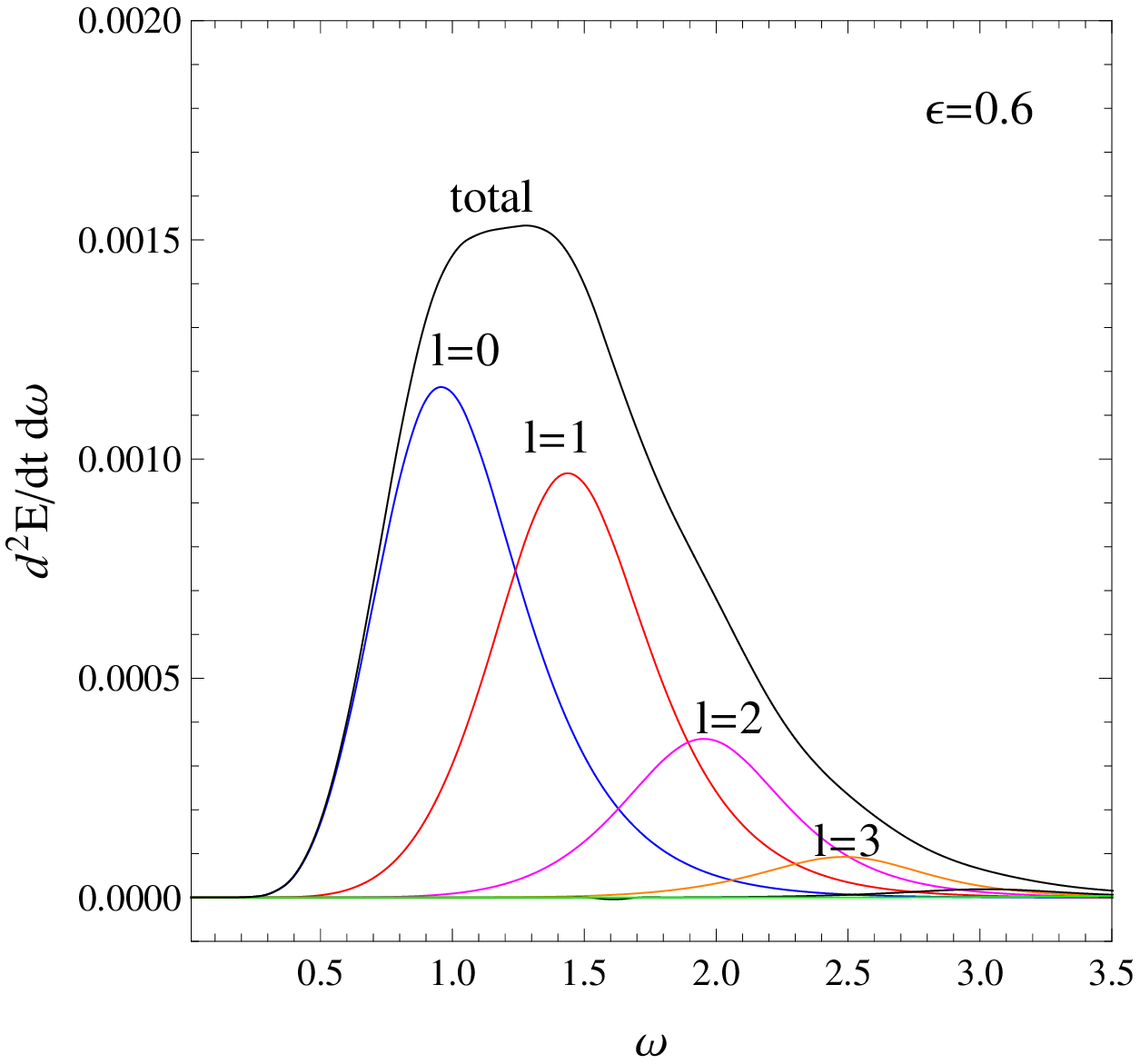} }
\caption{Energy emission rate for scalar field in the bulk for $n=2,Q=0.3,\Lambda=0.1$, (a) $\epsilon=0$, (b) $\epsilon=0.6$, with the first five modes $\ell=0,1,2,3,4,5$.} 
\label{FIGEERbulk1}
\end{figure}

In this subsection, we investigate the energy emission rates of the bulk scalar field {in} the higher-dimensional RN-dS {spacetime}. To compare with the results obtained in the preceding subsection, we deliberately choose parameters similar to those used in the brane case. The energy emission rates for each scalar field mode for six-dimensional RN-dS black holes were calculated and are presented in Fig.~\ref{FIGEERbulk1}. In these plots, the Bousso-Hawking temperature $T_{BH}$ is chosen, and the total emission rate is the total sum of the scalar field mode up to $\ell=5$. It is evident that the major contribution to the total emission rate is the result of the lower mode. In addition, we observe that the operation in the bulk scenario is similar to that of the brane case discussed earlier. For a minimally coupled scalar field with $\ell=0$, we find a non-vanishing value of the energy emission rate for a low-energy regime, as expected. The energy emission rates experience a suppression effect as the coupling parameter increases. However, the suppression from the coupling constant appears to have a lower effect on the bulk scenario. Finally, the energy emission rates of the bulk scalar (minimally and non-minimally coupled) are more suppressed than those of the brane scalar for the overall regime of $\omega$.

\begin{figure}[h]
\subfloat[]{ \includegraphics[width=0.45\textwidth]{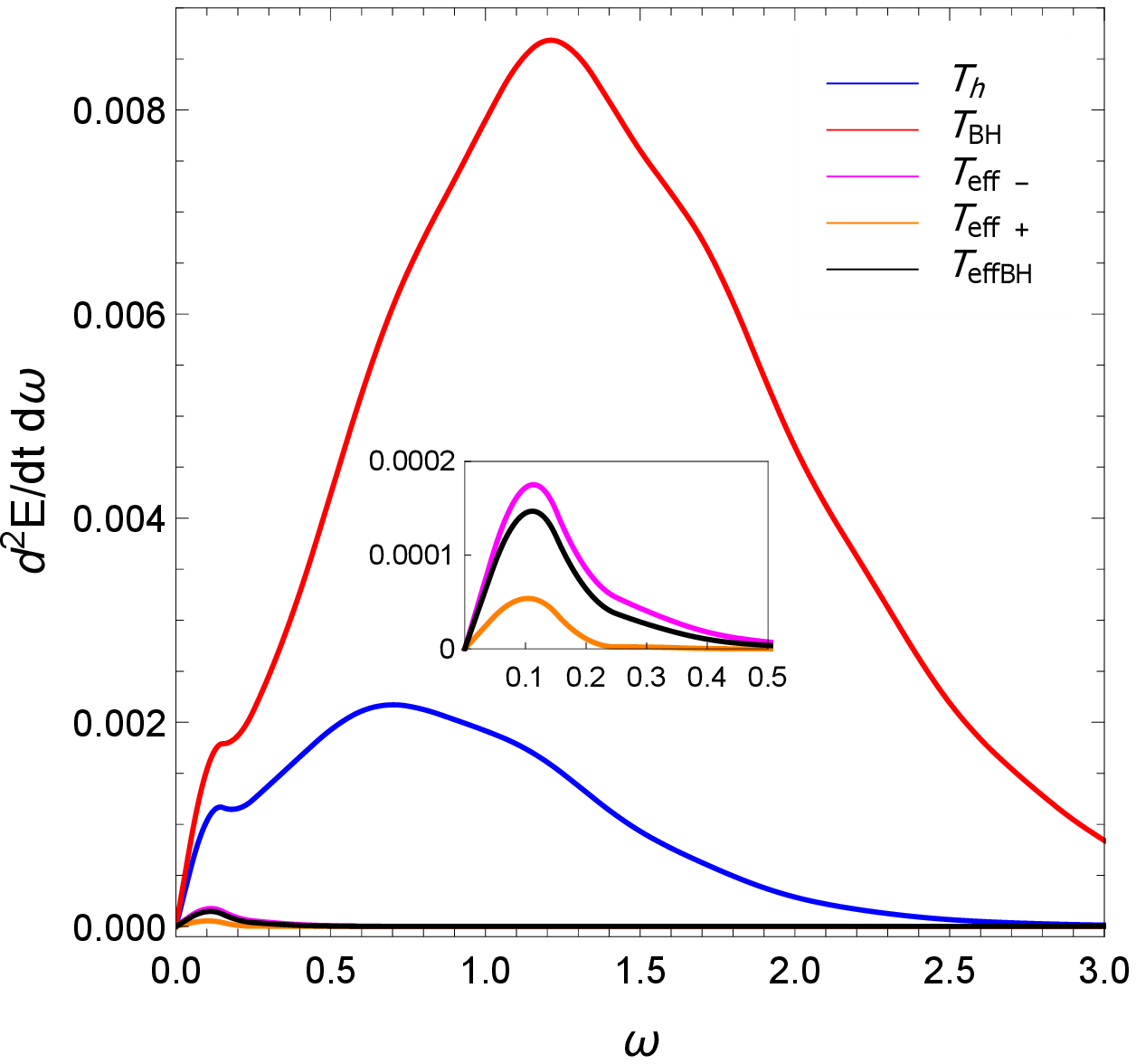} }
\subfloat[]{ \includegraphics[width=0.45\textwidth]{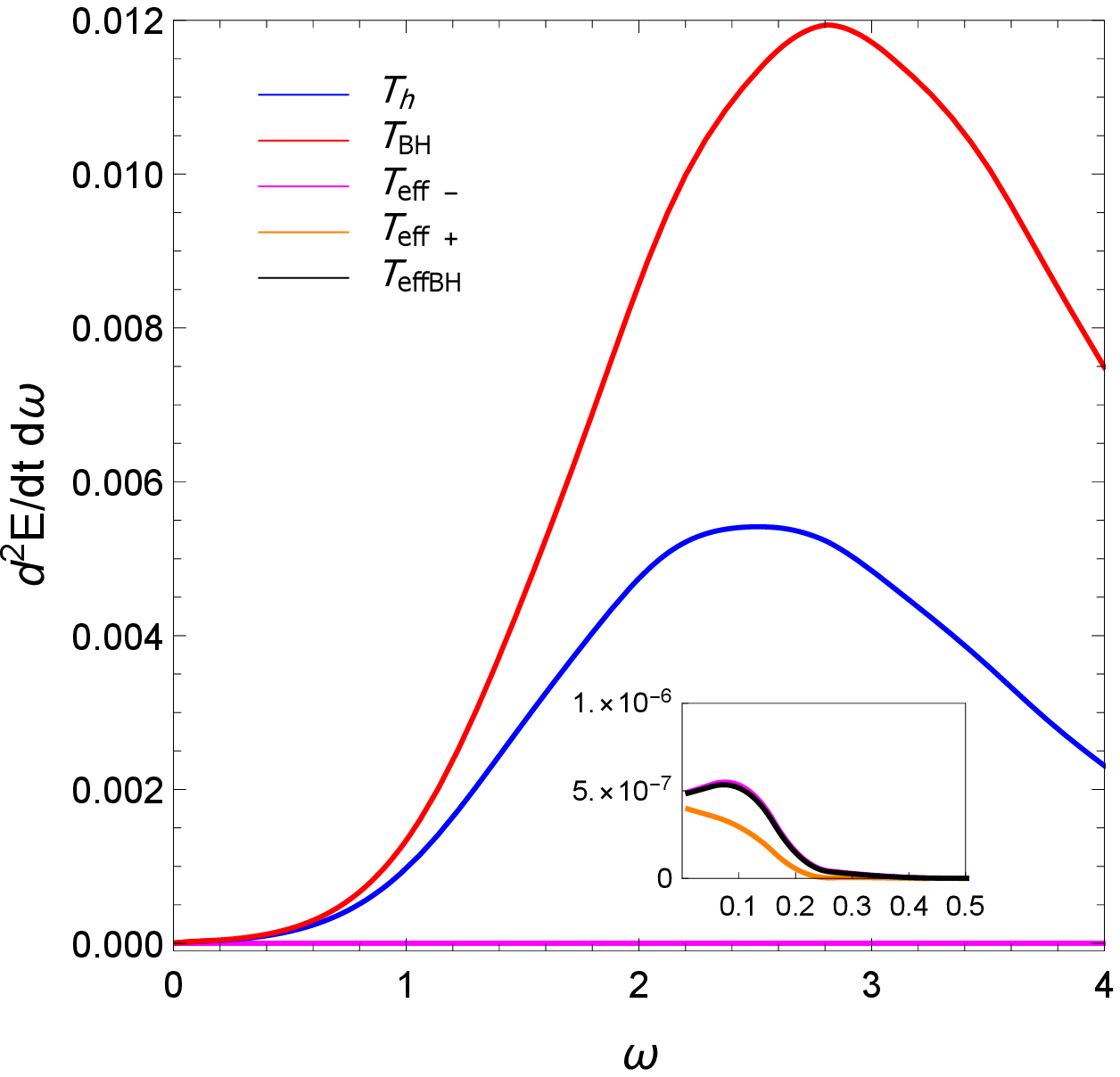} }
\caption{Comparisons of energy emission rate for scalar field in the bulk for each temperatures with $\epsilon=0.01,Q=0.1,\Lambda=1$, (a) $n=2$, (b) $n=4$.} 
\label{FIGEERbulk2}
\end{figure}

Fig.~\ref{FIGEERbulk2} depicts the energy emission rate of the non-minimally coupled scalar field in the bulk. The effects of the number of extra dimensions on the differential energy emission are presented in this figure. In contrast with the minimally coupled case, the energy emission rate becomes zero in the low-frequency regime. Similarly, the $T_h$ and $T_{BH}$ curves are higher than the effective temperature curves by approximately one order of magnitude. This is because  $T_{eff-},T_{eff+}$, and $T_{effBH}$ are comparatively smaller than the traditional and normalized black hole temperatures. As the number of extra dimensions increases, the $T_h$ and $T_{BH}$ emission curves are enhanced, whereas the rest become further suppressed. In comparison with the brane case (Fig.~\ref{FIGEERbrane2}), we find that the total energy emission rates for the scalar brane are generally larger than those for the bulk scalar field.

\begin{figure}[h]
\subfloat[]{ \includegraphics[width=0.45\textwidth]{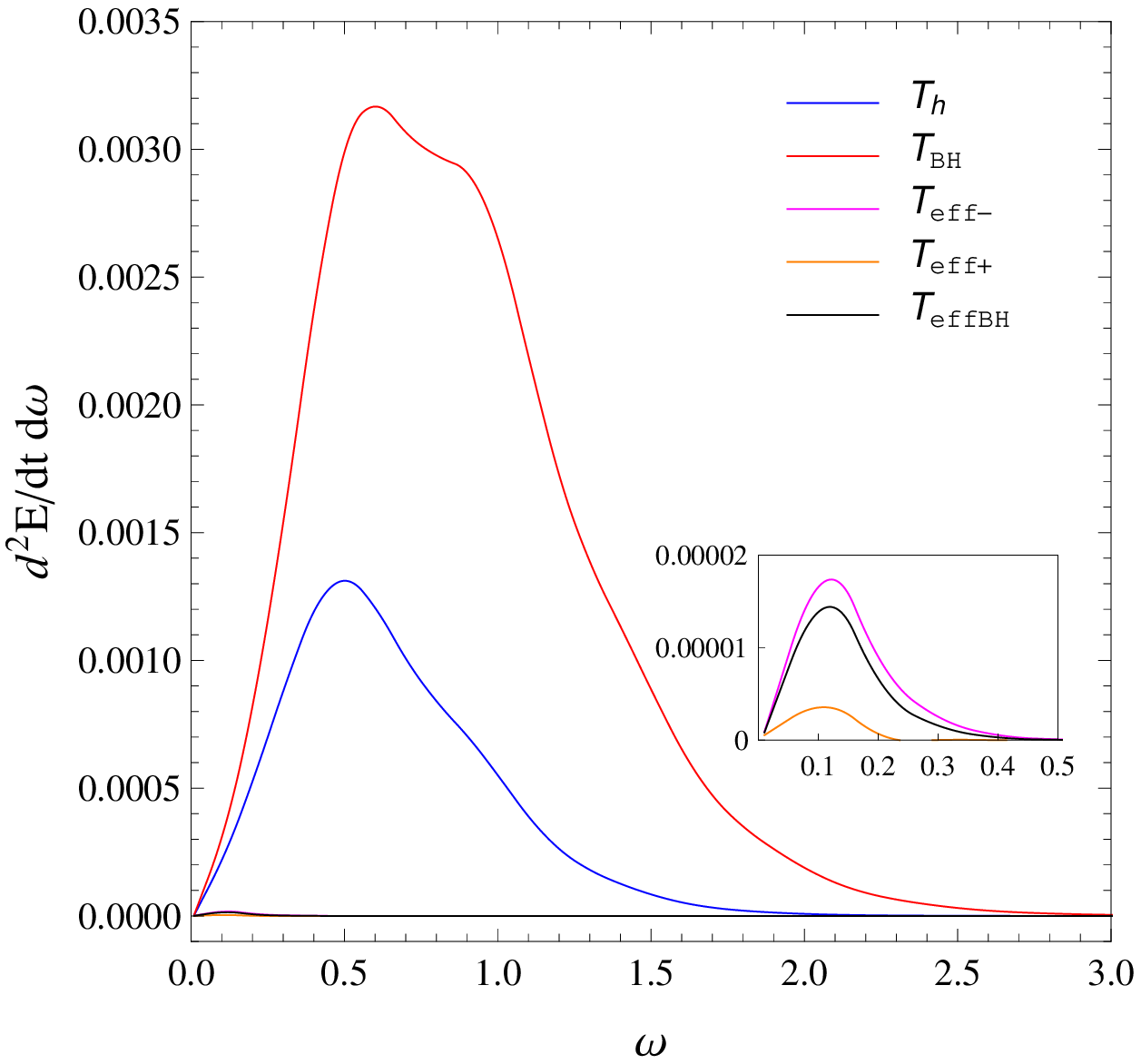} }
\subfloat[]{ \includegraphics[width=0.45\textwidth]{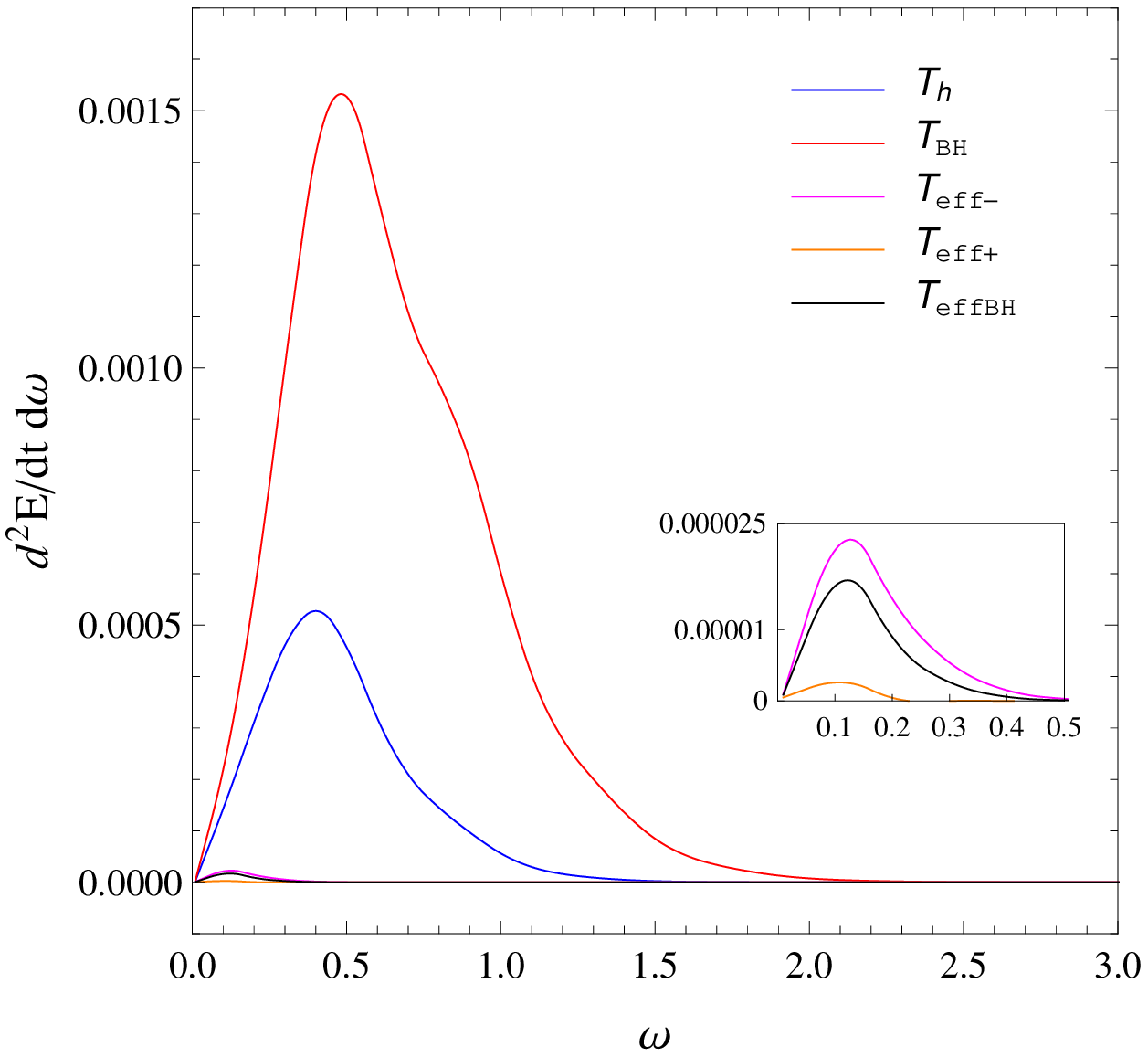} }
\caption{Comparisons of energy emission rate for scalar field in the bulk for each temperatures with $\epsilon=0.1,n=1,\Lambda=0.25$, (a) $Q=0.1$, (b) $Q=0.5$.} 
\label{FIGEERbulk3}
\end{figure}

The dependence of the charge of the black hole on the differential energy emission is depicted in Fig.~\ref{FIGEERbulk3}. Bell-shaped curves are observed in this case. The $T_h, T_{BH}$, and $T_{eff+}$ curves are suppressed by the increasing value of $Q$. Enhancements occur particularly for the $T_{eff-}$ and $T_{effBH}$ curves when the charge of the black hole increases. In both cases, $Q=0.1,0.5$, the differential energy emissions appear to favor the low-energy regime. These curves are found to be comparatively smaller than those of the scalar field on the brane displayed in Fig.~\ref{FIGEERbulk3}.

\begin{figure}[h]
\subfloat[]{ \includegraphics[width=0.45\textwidth]{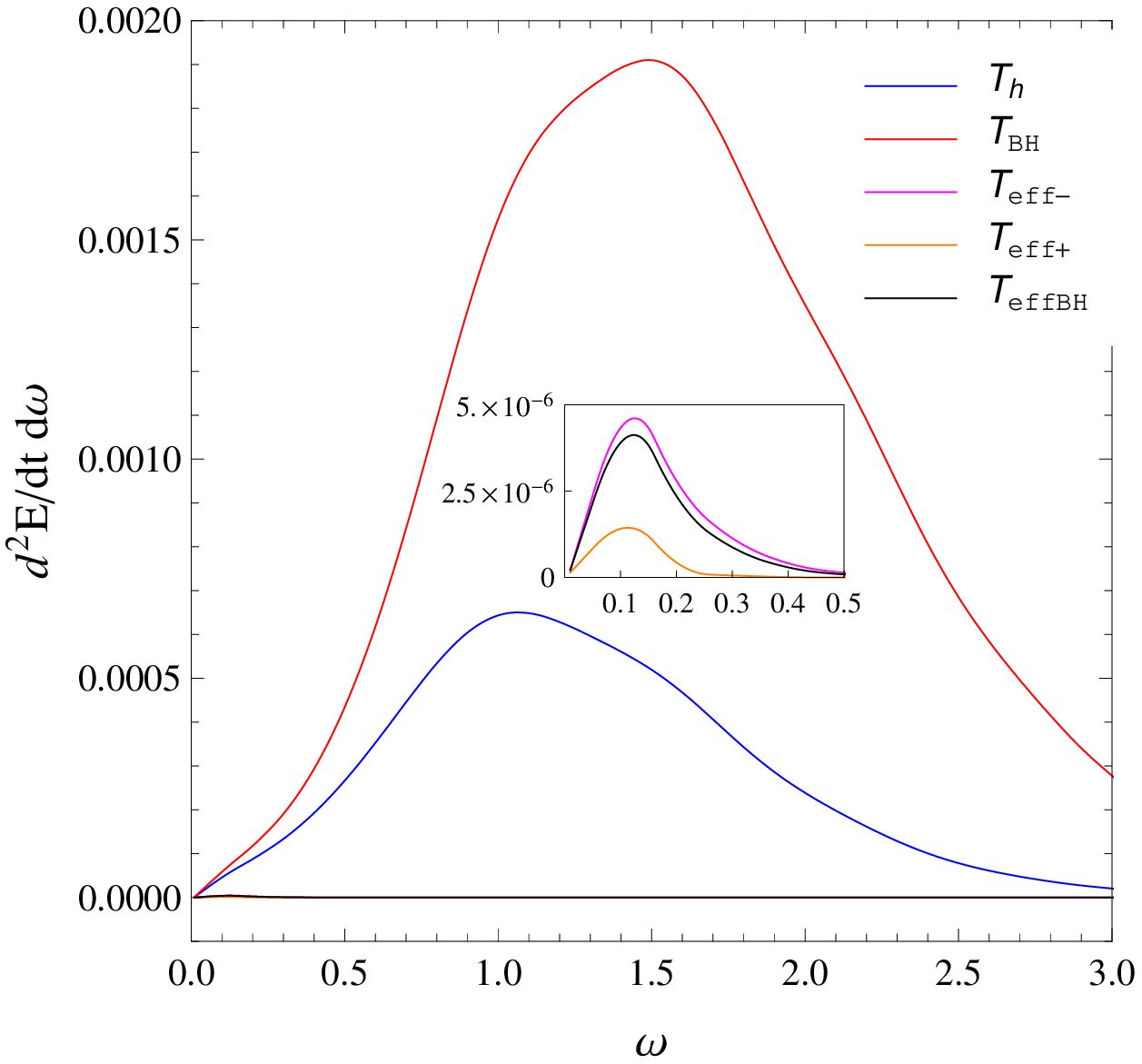} }
\subfloat[]{ \includegraphics[width=0.45\textwidth]{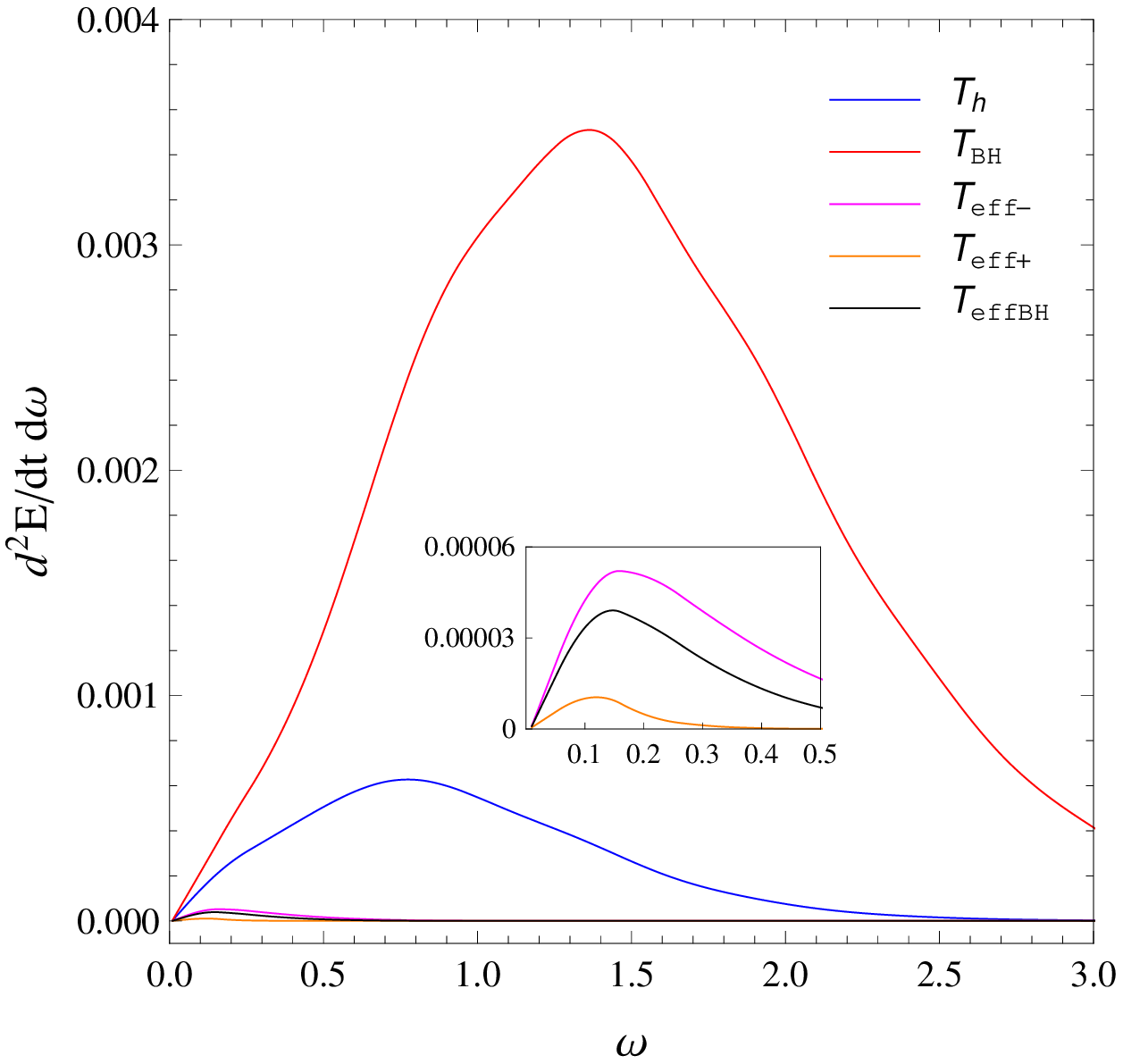} }
\caption{Comparisons of energy emission rate for scalar field in the bulk for each temperatures with $\epsilon=0.05,n=3,Q=0.5$, (a) $\Lambda=1$, (b) $\Lambda=2$.} 
\label{FIGEERbulk4}
\end{figure}

Fig.~\ref{FIGEERbulk4} depicts the energy emission rates of non-minimally bulk scalar field {in} the seven-dimensional RN-dS {background} with two specific values of the cosmological constant, that is, $\Lambda = 1,2$. In general, the energy emission curves are similar to those on the brane scalar field. Suppression occurs only in the $T_h$ curves when the cosmological constant increases. The other curves are all enhanced by the increasing value of $\Lambda$. When compared with the brane scalar case (FIG~\ref{FIGEERbrane4}), the energy emission curves for the scalar field in the bulk are further suppressed.

Notwithstanding the fact that the energy emission rates in the bulk do not differ significantly from those in the brane scenario, the bulk energy emission curves are found to be further suppressed in the overall regime of energy than those in the brane scenario. Moreover, in both the brane and bulk cases, the energy emission rates occur in the low- to intermediate-frequency regimes.

\section{Bulk-over-brane emission ratio}\label{sect:Bulkoverbrane}

In this section, we investigate the effect of various temperatures on the ratio of the total energy emitted by higher-dimensional RN-dS black holes in the bulk to that of the brane. We investigate the effect of the model parameters, particularly $Q$, $\Lambda$, and $\epsilon$, on the bulk-over-brane energy emission. To compute the total energy emitted in the bulk and on the brane, equation (\ref{emittbrane}) is numerically integrated over the entire range of frequencies $\omega$.

In Tables~\ref{tab:tab1} and \ref{tab:tab2}, we plot the bulk/brane emission ratios for five different temperatures of the black hole versus $Q=0.1,0.3,0.5,0.7$ with $\Lambda=1$. Note that $Q$ is chosen such that the temperatures of the black hole are finite. First, it is evident that brane emissions dominate the signal from the bulk. The emission ratios can be categorized into two types: those that decrease with $Q$ and those that increase with $Q$. The temperatures $T_h$ and $T_{BH}$ belong to the first group. We observe that the bulk-over-brane emissivities decrease when $Q$ increases. The results from Section~\ref{sect:EER} imply that the energy emission curves $T_h$ and $T_{BH}$ are suppressed with an increase in the charge. Therefore, the emission signals from the bulk gradually become less significant when compared with those from the brane. The second group consists of ratios from $T_{eff-},T_{eff+}$, and $T_{effBH}$. For $T_{eff+}$, the energy emission rates of both the brane and bulk scalar fields are suppressed with an increase in $Q$. In addition, the emission rates of the brane scalar field are less suppressed when compared with those of the bulk one, thereby enhancing the bulk/brane ratio when $Q$ increases. In contrast, when the energy emission rates are enhanced for both the $T_{eff-}$ and $T_{effBH}$ curves, it can be implied that the increase in the bulk emission signal when $Q$ increases is smaller than that in the brane case. Moreover, for the $T_{eff+}$ and $T_{effBH}$ cases, the ratios change significantly in the high-$Q$ regime. This is because both temperatures approach certain points at which they become infinite when $Q$ increases. This observation is depicted in Fig.~\ref{FIGtemp3}. Finally, when $\epsilon$ increases, the bulk-over-brane emission ratios are further suppressed throughout the entire range of parameters investigated.

\setlength{\tabcolsep}{15pt}
\begin{table}[h]

\centering

\begin{tabular}{|c|c|c|c|c|}
\hline
Temperature  & $Q=0.1$  & $Q=0.3$ & $Q=0.5$ & $Q=0.7$\\  \hline

$T_h$ & 0.301549  & 0.273211  & 0.221700 & 0.163230 \\
 
\hline

$T_{BH}$ & 0.501751  & 0.452067 & 0.352243 & 0.218765  \\
 
\hline

$T_{eff-}$ & 0.133887 & 0.135955 & 0.142676 & 0.188552   \\
 
\hline

$T_{eff+}$ & 0.125338 & 0.125657 & 0.126203 & 0.126572   \\
 
\hline

$T_{effBH}$ & 0.131824 & 0.133242 & 0.137299 & 0.154225  \\
 
\hline

\end{tabular}
\caption{Bulk-over-brane emission ratio for $\epsilon=0,n=2$ and $\Lambda=1$.}
\label{tab:tab1}
\end{table}

\setlength{\tabcolsep}{15pt}
\begin{table}[h]

\centering

\begin{tabular}{|c|c|c|c|c|}
\hline
Temperature  & $Q=0.1$  & $Q=0.3$ & $Q=0.5$ & $Q=0.7$\\  \hline

$T_h$ & 0.159281  & 0.129055  & 0.071677 & 0.013403 \\
 
\hline

$T_{BH}$ & 0.380857  & 0.326871 & 0.213148 & 0.059079  \\
 
\hline

$T_{eff-}$ & 0.003263 & 0.003635 & 0.005352 & 0.031760   \\
 
\hline

$T_{eff+}$ & 0.002555 & 0.002634 & 0.002819 & 0.003211   \\
 
\hline

$T_{effBH}$ & 0.002931 & 0.003113 & 0.003743 & 0.008527  \\
 
\hline

\end{tabular}
\caption{Bulk-over-brane emission ratio for $\epsilon=1,n=2$ and $\Lambda=1$.}
\label{tab:tab2}
\end{table}

\setlength{\tabcolsep}{15pt}
\begin{table}[h]

\centering

\begin{tabular}{|c|c|c|c|c|}

\hline
Temperature  & $\Lambda=2$  & $\Lambda=3$ & $\Lambda=4$ & $\Lambda=5$\\  \hline

$T_h$   & 0.401939  & 0.529195 & 0.646310 & 0.867203  \\
 
\hline

$T_{BH}$  & 0.793922 & 1.218580 & 1.932757 & 3.314220 \\
 
\hline

$T_{eff-}$  & 0.322934 & 0.536059 & 0.876612 & 2.201602  \\
 
\hline

$T_{eff+}$  & 0.299156 & 0.507383 & 0.728516 & 0.925172 \\
 
\hline

$T_{effBH}$  & 0.311829 & 0.500666 & 0.654044 & 0.882097 \\
 
\hline

\end{tabular}
\caption{Bulk-over-brane emission ratio for $\epsilon=0,n=2$ and $Q=0.1$.}
\label{tab:tab3}
\end{table}

\setlength{\tabcolsep}{15pt}
\begin{table}[h]

\centering

\begin{tabular}{|c|c|c|c|c|}

\hline
Temperature  & $\Lambda=2$  & $\Lambda=3$ & $\Lambda=4$ & $\Lambda=5$\\  \hline

$T_h$  & 0.086116  & 0.050423 & 0.030787 & 0.017883 \\
 
\hline

$T_{BH}$  & 0.510153 & 0.798298 & 1.445196 & 2.538121 \\
 
\hline

$T_{eff-}$  & 0.011776 & 0.056609 & 0.304208 & 1.602842  \\
 
\hline

$T_{eff+}$  & 0.003782 & 0.004598 & 0.005125 & 0.005418  \\
 
\hline

$T_{effBH}$  & 0.005529 & 0.008871 & 0.012074 & 0.012514 \\
 
\hline

\end{tabular}
\caption{Bulk-over-brane emission ratio for $\epsilon=1,n=2$ and $Q=0.1$.}
\label{tab:tab4}
\end{table}

Next, we investigate the effect of the cosmological constant on the bulk-over-brane emission ratios. For six-dimensional RN-dS black holes with $Q=0.1$, Tables~\ref{tab:tab3} and \ref{tab:tab4} present the total emission ratios versus $\Lambda$ for each temperature of the black hole. Similarly, the cosmological constant spans the range over which the temperature of the black hole is finite and positive. For the minimally coupled scalar field, it is evident that the bulk/brane ratio increase when $\Lambda$ increases. In addition, for most cases explored in this study, the total energy emissions from the brane dominate the bulk emissions. However, the latter becomes moderately more significant when $\Lambda$ increases. Notably, when $\Lambda$ increases sufficiently, the total emission ratios exceed unity, as demonstrated for $T_{BH}$ and $T_{eff-}$. This means that the emissions from the bulk channel overcome the signal from the brane channel. The dominance of the bulk emission signal over the brane emission is also observed in the higher-dimensional neutral dS black hole \cite{Kanti:2017ubd,Pappas:2016ovo}. When the coupling constant is nonzero, the total emission ratio is suppressed for the entire range of parameters. In general, the bulk-over-brane ratios are similar to those in the minimally coupled case. The exception occurs only for the traditional $T_{h}$. The bulk/brane emission ratios increase when $\Lambda$ approaches the maximum allowed value. This phenomenon occurs because the energy emission curves shift toward the high-frequency regime when $\epsilon$ increases, leading to the absence of the low-energy emission mode. However, the contribution of $T_h$ toward the energy emission curve in the high-energy regime is significantly small, because $T_h$ retains its maximum in the low-energy region \cite{Kanti:2017ubd}. Therefore, the suppression of the bulk/brane emission ratios when $\Lambda$ increases implies that the bulk emission signal is affected significantly more than the brane signal for the $T_h$ case.

\section{Conclusions}\label{sect:conclude}

In this study, we investigated the thermodynamics of higher-dimensional RN-dS black holes. In particular, we considered the effects of the model parameters on five different temperatures of the black hole: traditional temperature $T_h$; normalized (Bousso-Hawking) temperature $T_{BH}$; and three effective temperatures, namely, $T_{eff-},T_{eff+}$, and $T_{effBH}$. We first explored five temperatures under the effect of the cosmological constant $\Lambda$. In the limit where the cosmological constant vanishes, $T_{BH}$ was identical to $T_{h}$, as expected. The three effective temperatures were all zero, which indicated the invalidity of the effective temperature formulae in the absence of $\Lambda$. When $\Lambda$ approached its maximum allowed value, where the two horizons coincide, $T_h$ and $T_{eff-}$ gradually reached a nonzero value, whereas the rest of the temperatures tended to zero. Thereafter, the temperatures were studied under the effect of the number $n$ of the extra spacelike dimensions. Both $T_h$ and $T_{BH}$ monotonically increased with $n$, whereas the effective temperatures approached a nonzero (yet small) constant. The dependence of the five different temperatures on the charge of the black hole was discussed subsequently. The surface-gravity-based temperature $T_{h}$, Bousso-Hawking temperature $T_{BH}$, and $T_{eff+}$ decreased with $Q$ before reaching zero in the same limit. In contrast, the other two effective temperatures experienced infinite jumps in their values when the charge of the black hole approached its maximum allowed value. For the entire range of parameters investigated, the normalized temperature was dominant over the other definitions of the temperature of the black hole.

We also studied the scalar propagation on the brane and in the bulk. The scalar field was assumed to have non-minimal coupling to the curvature constant. The scalar curvature term was effectively the same as the mass term in the scalar equation of motion. We derived and numerically computed the transmission probability amplitude or greybody factor. We then investigated the effect of the model parameters, namely, coupling parameters $\epsilon, n, \ell, Q$, and $\Lambda$, on the greybody factors, both on the brane and in the bulk. Under the effects of these parameters, greybody factors share several common features on the brane and in the bulk. The scalar transmission amplitudes were enhanced with an increase in $Q,\Lambda$, whereas the greybody factors were suppressed with an increase in $\epsilon, n$, and $\ell$. Particularly, the suppression effect of the coupling parameters on the greybody factor in the bulk was softer than that on the brane. A notable difference was observed between the greybody factor with $\epsilon=0$ and that with $\epsilon\neq 0$ in the lowest dominant mode $\ell=0$ for both the brane and bulk, whereas it tended to a nonzero value for the former, it became zero for the latter when $\omega \to 0$.  

Next, we calculated the power spectra of energy emission based on the five aforementioned temperatures for a (non-)minimally coupled scalar field propagating on the brane and in the bulk. For both the brane and bulk scenarios, we found that the major contribution to the power spectra resulted from the first few lowest modes $\ell$. When $\epsilon$ took nonzero values, we noticed a suppression in the energy emission rates. Thereafter, we explored the dependences of $n, Q$, and $\Lambda$ on the power spectra. 
The nature of the emission curves was generally dictated by the temperature. When the number of extra dimensions increased, the emission rates for $T_h$ and $T_{BH}$ were enhanced, whereas those of the other effective temperatures were suppressed. The charge of the black hole affected the energy emission curves such that the $T_h, T_{BH}$, and $T_{eff+}$ curves were suppressed when $Q$ increased, whereas the $T_{eff-}$ and $T_{effBH}$ curves were enhanced. Moreover, only the $T_h$ curve was suppressed, whereas the other curves were enhanced when $\Lambda$ increased. For the entire range of parameters $n, Q$, and $\Lambda$ that we investigated, we observed that the emission rates for the traditional $T_h$ and Bousso-Hawking $T_{BH}$ were generally dominant over the effective temperature curves. In addition, we found that, for the same set of parameters, the energy emission rates of the scalar field on the brane were typically higher than those in the bulk.

Section~\ref{sect:Bulkoverbrane} presents a comparison of the total energy emissions along the brane and bulk channels. The bulk-over-brane emission ratios for the five temperatures were calculated while varying the various parameters. When the charge of the black hole increased toward the maximum allowed value, the total energy emission in the bulk channel became increasingly significant for $T_h$ and $T_{BH}$. For the effective temperatures, however, the emission ratios increased slightly when $Q$ increased. We also noticed an abrupt change in the values of the ratios when $Q$ approached the maximum value for the $T_{eff-}$ and $T_{effBH}$ cases. This reflected infinite jumps in the $T_{eff-}$ and $T_{effBH}$ curves, as discussed earlier. In general, the results of the total energy emissions confirmed the dominance of the scalar brane signal over the bulk signal. Nevertheless, when $\Lambda$ approached the maximum allowed value, the emission ratios were possibly larger than unity. Therefore, the bulk emission dominated over the emission channel at temperatures $T_{BH}$ and $T_{eff-}$. In general, the total energy emission ratios were suppressed when the coupling parameter was turned on, as expected. Note that we herein considered on black holes with extra dimensions. However, our observable universe, at macroscopic scales, has four dimensions, so the gravitational effects of extra-large spatial dimensions are not observed in our universe. Instead of large spatial dimensions, extra dimensions might be hidden by some short-scale mechanism such as compactification. Then, considering effects of the extra dimensions can be relevant in such a short range. Hence, our results might be important for phenomena involving small black holes with sizes comparable to the size of the extra dimensions. In this case, the gravitational effects owing to the extra dimensions can be detected by the radiation the black hole emits such as its Hawking radiation.

\acknowledgments
This work was supported by the National Research Foundation of Korea (NRF) grant funded by the Korea government (MSIT) (NRF-2018R1C1B6004349) and the Dongguk University Research Fund of 2021. We appreciate the APCTP for its hospitality during completion of this work.

\end{document}